\documentclass{aastex6}

\usepackage{colortbl}
\usepackage{graphicx}
\definecolor{LRed}{rgb}{1,.8,.8}
\begin{document}


\shorttitle{Segue 3 } 
 \shortauthors{Hughes et al.}

\title{A Multi-Wavelength Study of the Segue 3 Cluster}



\author{Joanne Hughes\altaffilmark{1}}
\affil{Physics Department
Seattle University \\
901 12th Ave.\\
Seattle, WA  98122, USA}

\author{ Brianna Lacy\altaffilmark{2}, Charli Sakari, George Wallerstein, Christoper Evan Davis}
\affil{Astronomy Department\\
 University of Washington\\ Box 351580\\ Seattle, WA 98195-1580, USA}

\author{Spencer Schiefelbein, Olivia Corrin, Hanah Joudi, Donna Le, and Rose Marie Haynes}
\affil{Physics Department
Seattle University \\
901 12th Ave.\\
Seattle, WA  98122, USA}



\altaffiltext{1}{jhughes@seattleu.edu}
\altaffiltext{2}{Current Address: Department of Astrophysical Sciences, 4 Ivy Lane, Princeton University, Princeton, NJ 08544}

\begin{abstract} We present new SDSS and Washington photometry of the young, outer-halo stellar system, Segue~3. Combined with archival VI-observations, our most consistent results yield: $Z=0.006 \pm 0.001$, $\log(Age)=9.42 \pm 0.08$, $(m-M)_0=17.35 \pm 0.08$, $E(B-V)=0.09\pm 0.01$, with a high binary fraction of $0.39\pm 0.05$, using the Padova models. We confirm that mass-segregation has occurred, supporting the hypothesis that this cluster is being tidally disrupted.  A 3-parameter King model yields a cluster radius of $r_{cl}=0.\degr017\pm 0.\degr007$, a core radius of $r_{c}=0.\degr003\pm 0.\degr001$, and a tidal radius of $r_t=0.\degr04 \pm 0.\degr02$. A comparison of Padova and Dartmouth model-grids indicates that the cluster is not significantly $\alpha$-enhanced, 
with a mean $\mathrm{[Fe/H]}=-0.55^{+0.15}_{-0.12}$~dex, and a population age of only $2.6\pm 0.4$~Gyr. We rule out a statistically-significant age-spread at the main-sequence turn-off \rm because of a narrow sub-giant branch, \rm and discuss the role of stellar rotation and cluster age, using Dartmouth and Geneva models: \rm approximately $70\%$ of the Seg~3 stars at or below the main-sequence turn-off have enhanced rotation. \rm Our results for Segue~3 indicate that it is younger and more metal-rich than all previous studies have reported to-date. From colors involving Washington-C  and SDSS-u filters, we identify several giants and a possible blue-straggler for future follow-up spectroscopic studies, and we produce spectral energy distributions of  previously known members and potential Segue 3 sources with Washington ($CT_1$), Sloan ($ugri$), and $VI$-filters.   Segue~3 shares the characteristics of unusual stellar systems which have likely been stripped from external dwarf galaxies as they are being accreted by the Milky Way, or that have been formed during such an event. Its youth, metallicity and location are all inconsistent with Segue 3 being a cluster native to the Milky Way.

\end{abstract}

\section{Introduction} \label{sec:intro}

A form of ``stellar archaeology" traces the formation of the Milky Way (MW) using the dense globular clusters (GCs) as test particles. However, it has become obvious that globular cluster populations are far more chemically-diverse than we assumed a few decades ago, and are not simple single-generation star clusters \citep[and references therein]{gra12}. The Milky Way (MW) contains at least 150 GCs, and there appears to be a difference between the inner and outer halo populations \citep{van13}.

Segue 3 was first discovered by \citet{bel10} in the Sloan Digital Sky Survey (SDSS) ($\alpha =21^h 21^m 31^s$, $ \delta = +19\arcdeg 07\arcmin 02\arcsec $ J2000, $l=69.\arcdeg4$, $b=-21.\arcdeg27$), and was identified as an ultra-faint star cluster with a half-light radius, $r_h=0.065\arcmin \pm 0.1\arcmin$. The discovery paper detailed KPNO 4-m g- and r-photometry used to derive Seg~3's structure, employing an M92-like template isochrone. \citet[hereafter B10]{bel10} found $(m-M)_0=16.3$ and $\mathrm{[Fe/H]}=-2.3$, which indicated that Seg~3 is a cluster similar to Koposov 1 \& 2 \citep{kop07}, and tentatively linked Seg~3 with the structure of the Hercules-Aquila Cloud. 

\citet[hereafter, F11]{fad11} used Keck/DEIMOS spectroscopy and Magellan/IMACS g and r-band imaging of Seg~3, coupled with maximum likelihood methods,  to analyze the structure of the star cluster. F11 found a smaller $r_h$ of $26\arcsec \pm 5\arcsec$, with an age of $12^{+1.5}_{-0.4}$~Gyr and $\mathrm{[Fe/H]}=-1.7^{+0.07}_{-0.27}$. F11 identified 32 member-stars from spectroscopy and photometry, and placed 11 of the stars outside three of their half-light radii, finding no evidence of dark matter. F11 support  B10's conclusion that Seg~3 is an old, faint, sparse star cluster.  B10 note that the evolution of a system like Seg~3 proceeds with more massive objects collecting at the core of the cluster, with less massive objects forming a halo, as it disrupts. \citet{kim15} discuss the cluster Kim~1, mentioning Seg~3; they concluded that since F11 found radial a velocity offset between Seg~3 and the Hec-Aql Cloud, they were not likely to be connected.

In contrast, \citet[hereafter, O13]{ort13}  used deep Galileo (Telescopio Nazionale Galileo) B, V and I images of Segue 3 ($V\leq 25$) to determine an age of $\sim 3.2$~Gyr and $\mathrm{[Fe/H]}\sim -0.8$. The O13 result characterized Seg~3 as the youngest globular cluster in our Galaxy. Its likely youth may imply that Seg~3 is a captured object or a system formed during a capture of a gas-rich dwarf (O13; F11). Such gas-rich dwarf galaxies (e.g., WLM, SMC, LMC) may have donated clusters with properties similar to Palomar 1 \citep{sar07,sak11} and  Seg~3 to the Milky Way (MW). With a well-defined MSTO in their $V \; vs. \; (V-I)$ color-magnitude diagram (CMD), O13 found $(m-M)_0=17.32$, $d_\odot= 29.1$~kpc, with Galactic coordinates of $X = -13.0$, $Y = -6.1$, and $Z = -19.2$, making its Galactocentric distance $R_{GC} = 24.0$~kpc -- placing it among the unusual, outer-halo faint clusters. 

A study by \citet{pau14} identified Ko 1 \& 2 as \em open clusters \rm (OCs) of ages 5--7 Gyr, with $\mathrm{[Fe/H]}=-0.60$ and $[\alpha/Fe]=+0.2$, that could have been lost by the Sagittarius Dwarf Galaxy, and are part of the Sagittarius Stream. Their
conclusion was based on evidence that these clusters' luminosity functions (LFs) did not show significant mass-loss. O13's V-band LF shows  Seg~3's main-sequence is depleted above their completeness limit, indicating that it has undergone significant tidal-stripping, and should have been a more massive system in the past.

 \begin{figure}
\begin{minipage}[r]{0.7\textwidth}
\vspace{-4in}
\includegraphics[scale=0.825]{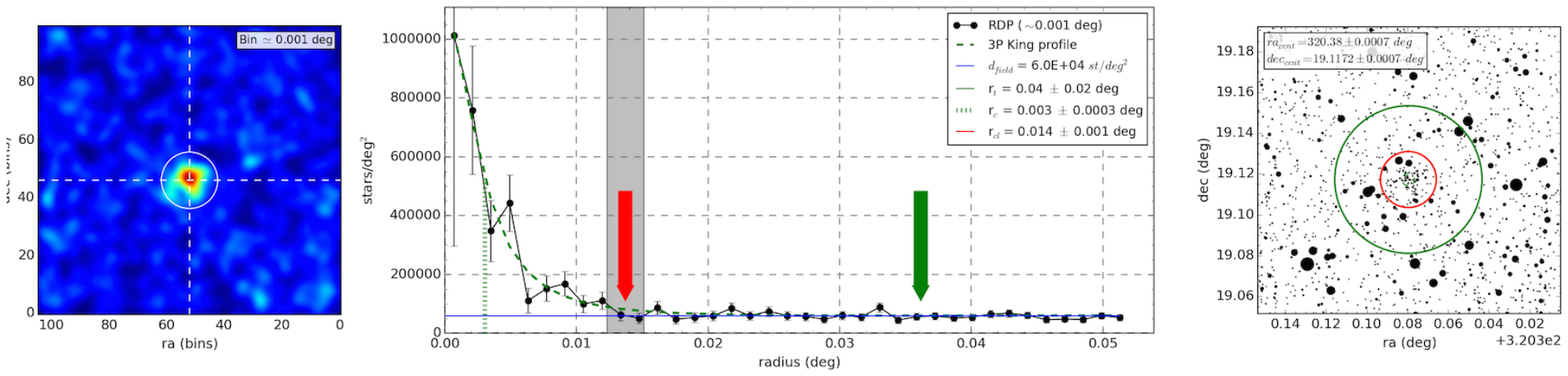}
\end{minipage}
\vspace{-4.0in}
\caption{
The Segue 3 cluster with the panels as follows. Left: the source-density plot, using data from O13. Center: the 3-parameter (3P) King model for the $V-I$ data. Right: finding chart for objects with the cluster and tidal radius shown.
The point-size represents the V-magnitude.
 }
\label{fig:f1}
\end{figure}
 \begin{figure}
\includegraphics[scale=0.6]{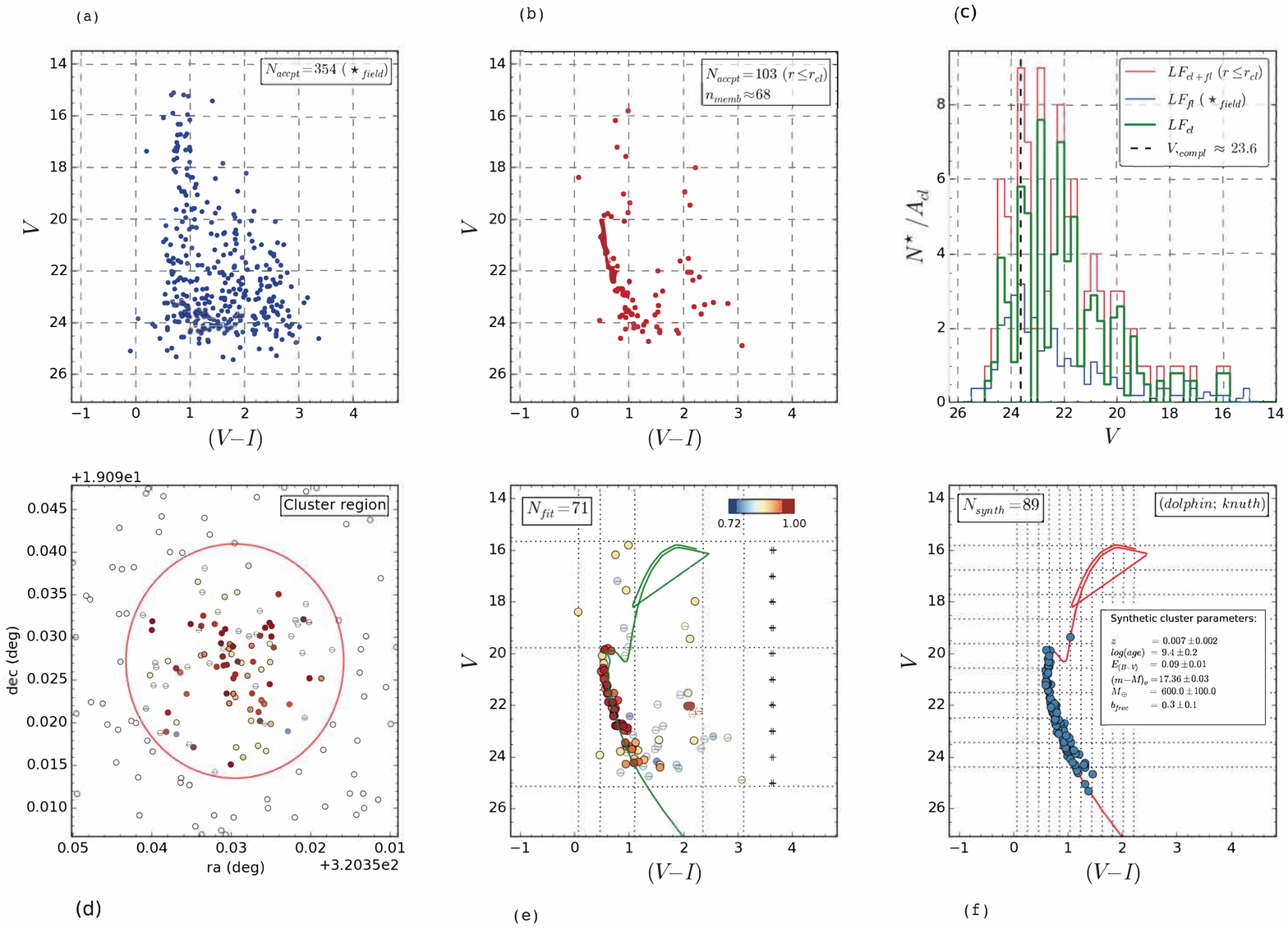}
\caption{
The Segue 3  (O13) data re-analyzed with the ASteCA code (Perren et al. 2015). \rm (a): \rm left: CMD for stars in the FOV which are statistically likely to be non-members. \rm (b) \rm CMD for stars likely to be members. \rm (c) \rm Luminosity functions (LFs) in the form of number of stars/cluster area ($N^* / A_{cl}$) vs. V-mag., for the samples: the red line represents all stars within $r_{cl}$, the blue line denotes the ``field" population, the green line shows the likely Seg~3 members, and the dotted line is the completeness limit in the V-band. \rm (d) \rm A finding chart for the objects within the cluster area (see Figure~1), color-coded for the likelihood of membership. \rm (e) \rm CMD of likely members with the best-fit Padova isochrone (O13's data tables did not list uncertainties). \rm (f) \rm The synthetic cluster produced by ASteCA from the Padova/PARSEC12 isochrones, using the best-fit parameters (inset). This single run results in: $Z=0.007 \pm 0.002$, $\log(Age)=9.4 \pm 0.2$, $(m-M)_0=17.36 \pm 0.03$, and a binary fraction of $0.3\pm 0.1$. The 3-parameter (3P) King model gives  a cluster radius of $r_{cl}=0.\degr014\pm 0.\degr001$, the core radius is $r_{c}=0.\degr003\pm 0.\degr0003$, and the tidal radius is $r_t=0.\degr04 \pm 0.\degr02$, yielding a total mass of $600\pm 100\; M_\odot$.
 }
\label{fig:f2}
\end{figure}

O13 argued that the main difference in their results and those of F11 was caused by an offset (of unknown origin) in the latter's photometry, and the inclusion or exclusion of a few sub-giant branch (SGB) stars. No red giant branch (RGB) objects have been conclusively identified as members of Seg~3 from past studies (B10, F11, O13), and no spectroscopic metallicities have been reported.  As the ``youngest globular cluster" (O13) in the MW, this is an important system, which merits further study. We attempt to reproduce the previous results in \S 2.1, using the archival O13 data.
In this paper, we study Seg~3 with Washington (\S 2.2) and SDSS (\S 2.3) filters, both to provide more wavelength coverage and to reduce the observational uncertainties in the age and metallicity. The $(C-T_1)$ and $(u-g)$ colors are around 2--3 times more sensitive than $V-I$ and $(g-r)$ \citep[depending on metallicity and age]{gei99,li08,hug14}. A discussion in \citet[and references therein]{hug14} compared previous papers that tested the most effective color-pairs in use for age and metallicity studies \citep[etc.]{li08,hol11}, noting that the theoretical colors were tested on relatively close and dense GCs.
 
To avoid user-bias as much as possible, we compared the results of fits to the Dartmouth models \citep{dot08,dot16} made with simple $\chi^2$-fitting routines for multiple colors, with more complex open-source codes that claim to simultaneously fit 7--9 parameters which are degenerate in color-magnitude  diagrams or color-color plots. We also chose 2 different codes, one which has been tested on OCs and one designed for GCs.
To compare directly with O13, who used the \citet{bre12} models, we employed the PARSEC \citep[stellar tracks and isochrones with the PAdova and TRieste Stellar Evolution Code] {bre12} and the (open-source) Automated Stellar Cluster Analysis (ASteCA) suite of modeling tools  \citep[ \S 2.1]{per15}. 

To better estimate the observational uncertainties, compared to the standard age-metallicity scale \citep{van13} for GCs, we used BASE-9 \citep{hip14,ste16,wag16a}. This is another Bayesian modeling code which fits star cluster basic parameters but it requires cluster-membership be assigned to stars in the region, and Segue 3 is a sparse stellar cluster in a crowded field. An advantage of the
 ASteCA suite of tools is that it contains  a ``Bayesian field star decontamination algorithm capable of assigning membership probabilities using photometric data alone." BASE-9 can be used for single-age and single-metallicity clusters \citep{wag17} but can also be set to model clusters that differ internally in helium abundance \citep{ste16,wag16b}, using the Y-enhanced Dartmouth isochrones (D08). 
 For completeness in considering stellar rotation as a free parameter, we also compared our data with the Geneva model database \citep{geo13}, which allows for a wider range of stellar rotation rates than the Dartmouth models. 
 
Without medium- to high-resolution spectroscopy,  we cannot confirm that Segue 3 is (or was, before it was so severely stripped: O13, F11, B10) a ``standard" GC, with the Na-O anti-correlation, denoting multiple-populations \citep{gra12}.
Helium-enhancement can
affect colors and  might be a valid discriminator between OCs and GCs. We also searched for similarities between Seg~3 and  the LMC 
\& SMC young massive clusters \citep[YMC, and references therein]{bas16}.
 
We detail our observations in \S 3. In \S 4, we discuss the spectral energy distributions and ensure all the photometry (UV--IR) can be calibrated to a uniform metallicity scale for cluster members and possible RGB stars, and one
likely blue straggler. In \S 5, we discuss the possibility that we are observing a spread in rotation rates instead of a large age-uncertainty at the turn-off. We address the age-metallicity relationship for galactic globular clusters in \S 6, and place Segue 3 in a group of unusual outer-halo systems that might be extra-galactic in origin.

\section{Method}

\subsection{Comparison of VI-data with the Padova Models}

We obtained the archival O13 data (noting that uncertainties are not available) and display the ASteCA fits in Figures 1 \& 2. Figure~1 shows the source density map, the 3-parameter King model fit, and the finding chart for objects scaled by V-mag. Figure~2 shows the $V,\; V-I$ ASteCA cleaning process, where 10 ``field regions" were defined around the cluster. For the O13 FOV, relaxing or tightening the cluster-membership criteria in the code  \citep{per15} selected $>50$ members in those filters. We show one run for the data, setting the visual extinction range to $0.05<E(B-V)<0.20$~mag., as we found that ASteCA can only reproduce the O13 results exactly by limiting the input interstellar extinction, and by forcing $Z<0.005$. The lack
of a well-defined RGB requires us to limit the extinction range; our tests added artificial stars to an assumed RGB to confirm this.
We limited the distance modulus to 15--20 magnitudes and let all other parameters range over the usual Padova/PARSEC12 model grid \citep{bre12}: we searched a $\log (Age)$ range from 6.0--10.13 and a metallicity range from Z=0.0001--0.015.  We ran the code in manual and automatic mode to test the stability of the fit to a cluster with known members (F11 \& O13) outside the apparent cluster radius. 

\rm We took the extinction values from the \citet[hereafter: SFD]{sch98} IRAS maps, noting that the MSTO magnitude/color is very sensitive to the assumed extinction, and can change the age and metallicity considerably in fitting isochrones. For the extinction corrections we assume the  relationships given in Eqns.[1]--[8]. \rm In the Washington filters, we use standard relationships from \citet{gei91} and \citet[hereafter: GS99]{gei99} for Washington filters, and those listed by \citep[and SFD]{yua13}. GS99 use $A_V = 3.2E(B-V)$; not setting $R_V=3.1$ does not transform into an appreciable difference with low E(B-V), but R-values might vary in different galaxy environments. Previous studies found these relationships would return photometric metallicities which compared well to spectroscopic measurements \citep{hug08,hug14}. Both O13 and F11 results agree with the SFD/IRAS maps: $E(B-V)\approx 0.1$, and extinction does not appear variable. 

\begin{eqnarray}
E(V-I)=1.24E(B-V);\\
E(C-T_1) = 1.97E(B-V);\\
E(T_1-T_2) = 0.69E(B-V);\\
M_{T_1}=T_1+ 0.58E(B-V) - (m-M)_V;\\
u_0=u-5.14E(B-V);\\
g_0=g-3.79E(B-V);\\
r_0=r-2.75E(B-V);\\
i_0=i-2.09E(B-V).
\end{eqnarray}

We allowed the cluster radius to vary within the limits set by B10, F11, and O13. The ASteCA ``radius assignment function" can be set to low/mid/high to determine how ``aggressively" the routine assigns the radius of the cluster. The \em low \rm option selects smaller radii, which is most useful when the FOV is heavily contaminated by field stars, and the \em high \rm option is used when to try and find every possible member belonging to the cluster. The example in Figure~2 uses the \em mid-\rm option. We set the Bayesian cleaning process to take 10 (a variable) field regions around the cluster  and found the membership probability of each object (Figure~2d \& 2e). The code can then find the likelihood of each synthetic star cluster generated \citep{dol02} and we show the  cluster members, color-coded for probability of membership (Figure~2e) with the best-fit isochrone \citep[and references therein]{per15}. There are some stars in the FOV, appearing to be on the RGB, which are not removed by the membership assignment, and one A-type star remains above the MSTO ($^\# 26$) -- possibly a blue straggler (BS) that was never rejected as a cluster member by the ASteCA-code, in any color combination.
 
The distance modulus settled down to $17.4\pm 0.5$ mag., when $E(B-V)$ was limited. Again, as noted by O13, it is sensitive to the inclusion of objects close to the (supposed) SGB and includes a few RGB/AGB stars which are not statistically excluded.  The limiting magnitude is calculated as $V\sim 23.6$ mag. by the code.  The single run shown in Figure~2 gives: $Z=0.007 \pm 0.002$, $\log(Age)=9.4 \pm 0.2$, $(m-M)_0=17.36 \pm 0.03$, and a binary fraction of $0.3\pm 0.1$. The 3-parameter (3P) King model returns a cluster radius of $r_{cl}=0.\degr014\pm 0.\degr001$, a core radius of $r_{c}=0.\degr003\pm 0.\degr0003$, and a tidal radius of $r_t=0.\degr04 \pm 0.\degr02$, yielding a total mass of $600\pm 100 \; M_\odot$. Thus, only limiting $E(B-V)$ reproduces the O13 results for the distance and size, but returns a more metal-rich and younger fit. \rm These results place Seg~3 firmly in the MW's outer halo, supporting the hypothesis that the cluster is disrupting:  the MS is underpopulated above the limiting magnitude, as reported in O13.  \rm There are $V_r$-confirmed cluster members outside both the $r_h$ (F11, O13) and  the average $r_{cl}$, determined by the code. The range of the tidal radius is quite uncertain (F11, O13). 

In Figure~2c, the contamination index (CI) is defined as:
\begin{equation}
 CI={d_{field}\over{n_{cl+fi}/A_{cl}}}={n_{fi}\over{n_{fi}+n_{cl}}},
 \end{equation}
 where $d_{field}$ is the field-star density in the cluster region, $A_{cl}$ is the area of the cluster, $n_{fi}$ is the number of field stars, and  $n_{cl}$ is the number of likely cluster members. For this example,  the O13 FOV has a value of $CI=0.34$; where a value $>0.5$ would indicate an equal number of field and cluster stars. \citet{per15} discuss the limitations of this code when dealing with a region which is heavily contaminated. The photometry is listed in O13 to $V\sim 25$ (no uncertainties), although ASteCA's routines calculate $V\approx 23.6$ as the completeness limit. ASteCA's analysis was tested on 400 MASSCLEAN-generated clusters \citep{pop09} by \citet{per15}, who also modeled 20 MW OCs, where Seg~3 is at the lower-mass end of the OC-sample they used. 

\subsection{Washington Filters}

In addition to studying the ASteCA error-analysis from \citet{per15}, we tested the code ourselves on Washington photometry of several GCs, most notably NGC 6397 and 47 Tuc, which were used as cluster-standards in \citet{hug07} for comparison with the massive, unusual GC, NGC 6388. We chose these 2 clusters since they are close to the values of [Fe/H] reported by F11 and O13 for Seg~3. NGC 6397 was modeled as $Z=0.0005 \pm 0.0001$, which becomes $\mathrm{[Fe/H]}=-1.92 \pm 0.11$ translated to the $\alpha$-enhanced scale from $\mathrm{[Fe/H]}=-1.6$ (solar-scaled). \citet{van13} reports  $\mathrm{[Fe/H]}=-1.99$. Also, $\log(Age)=10.1 \pm 0.05$, $E(B-V)=0.14 \pm 0.02$, $(m-M)_0=12.19 \pm 0.04$, and the binary fraction is found to be $0.30 \pm 0.09$. For 47 Tuc, the results were: $Z=0.0027 \pm 0.0002$ which is $\mathrm{[Fe/H]}=-0.82 \pm 0.06$ (solar-scaled), $\log(Age)=10.1 \pm 0.05$, $E(B-V)=0.04 \pm 0.02$, $(m-M)_0=13.33 \pm 0.07$. However, the binary fraction returned by the code was too high at  $0.50 \pm 0.06$; this GC has a broader MSTO and it is a very crowded field for ground-based telescopes, producing more blended stellar images than a sparse cluster. The ASteCA code was able to fit isochrones with a reasonable match to accepted literature-values for these GCs, which should bound the range in metallicities expected, and the OC data showed that the ages were acceptable for much younger systems.

\begin{figure}
\includegraphics[scale=0.6]{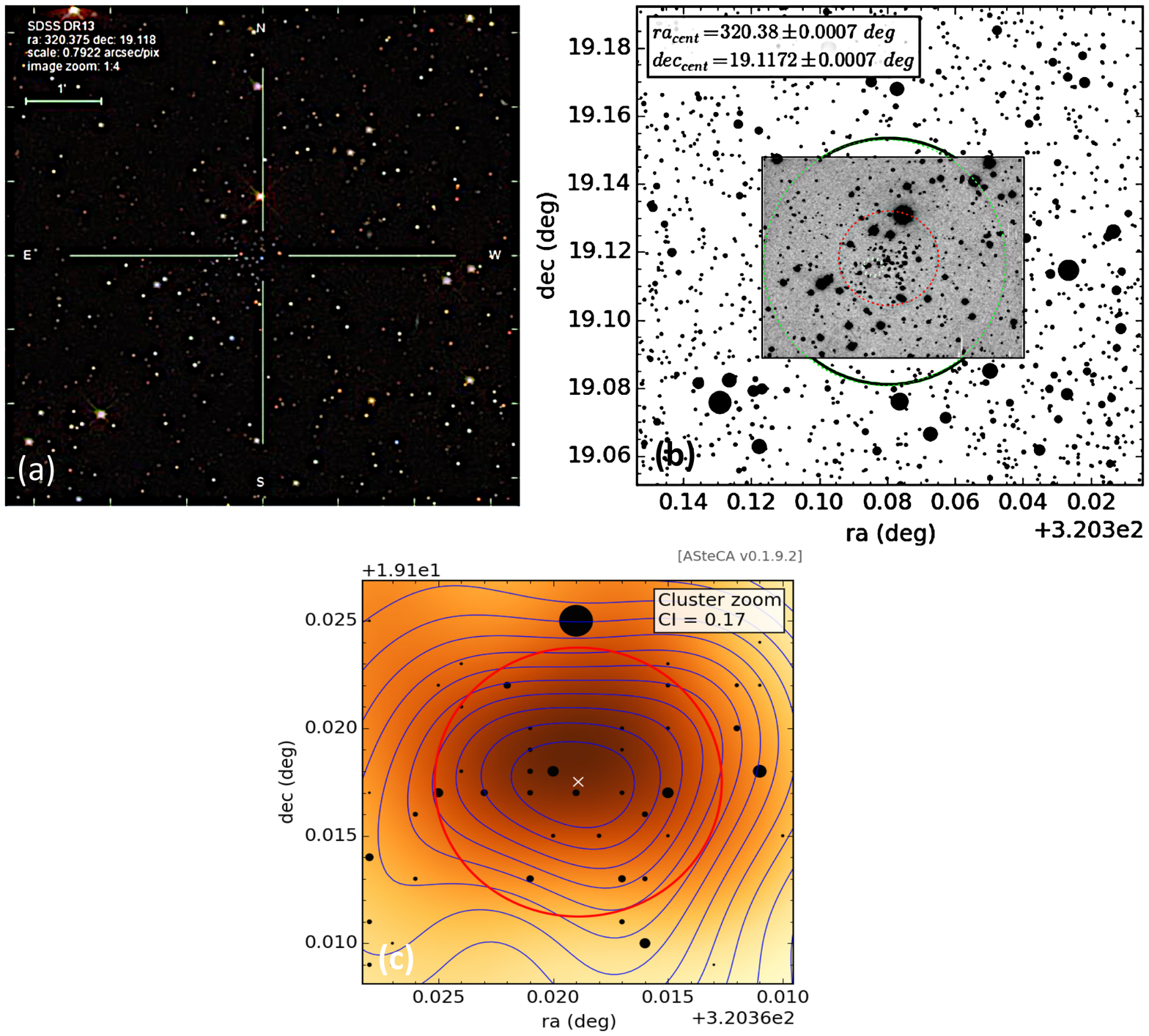}
\caption{\rm (a) \rm The SDSS DR13 image of the Segue 3 field. \rm (b) \rm The O13 V-filter finding chart (Figure 1-far right) is shown for the Segue 3 cluster with an overlay of the co-added, R-filter image (later converted to $T_1$ photometry) of the SPICam/ARCTIC combined FOV. The overlapping images result in an effective-FOV of 
$4.\arcmin 0 \times 3.\arcmin 5$.  The 3-parameter King model is shown for a fitted cluster radius of $r_{cl}=0.\degr017\pm 0.\degr007$ (red dashes), a core radius is $r_{c}=0.\degr003\pm 0.\degr001$ (pale green double-dashes), and the tidal radius is $r_t=0.\degr04 \pm 0.\degr02$ (green circle). The CI for the larger FOV for O13 is 0.34.
\rm (c) \rm The source density contours for the Washington filters, with CI=0.17, for our FOV. 
 }
\label{fig:f3}
\end{figure}

The Washington filters \citep{can76} $C$ and $T_1$ have advantages over other photometric systems due to the short integration times of the broadband filters, the metallicity-sensitivity, and the wealth of previous studies of galactic and extragalactic globular clusters \citep[GS99]{gei91}. A recent paper \citep{cum17} discussed the importance of the C-filter over the narrower SDSS-u, also examining F336W, for the study of multiple populations in GCs. Specifically, the paper concentrates on NGC 1851, on the RGB and SGB, and noting that the C-filter could be more effective at detecting multiple MSs. \citet{cum17}  also note that C and F336W can be affected by CN/CH variations. 
The original metallicity indicators were $(M-T_1)$ and $(C-M)$, with the latter color used most for metal-poor stars \citep{gei86,gei91}. Most previous extra-galactic studies used  the $(C-T_1)$-color \citep{gei90}: it is very sensitive to age and metallicity on the RGB (GS99).   For  $\mathrm{[Fe/H]}<-2.5$, only the Washington C-filter was found to be very sensitive to $\alpha$-enrichment \citep{hug14}, with its center at 3900\AA\ and a FWHM of 1100\AA\ \citep{can76}. The $(C-T_1)$ or C-[Kron-Cousins R], which is more-commonly used (GS99), does lose some metallicity resolution around $\mathrm{[Fe/H]}\approx -2$. However, testing of artificial stars with metallicities ranging from $-2.5<\mathrm{[Fe/H]}<-0.5$ (generated from the Dartmouth models in Hughes et al. 2014), showed that $(C-T_1)$ should have twice the sensitivity of $(V-I)$ in finding the metallicity of a star cluster with $\mathrm{[Fe/H]}\approx -1.8$, and improves for higher-metallicity systems.

 \begin{figure}
\includegraphics[scale=0.8]{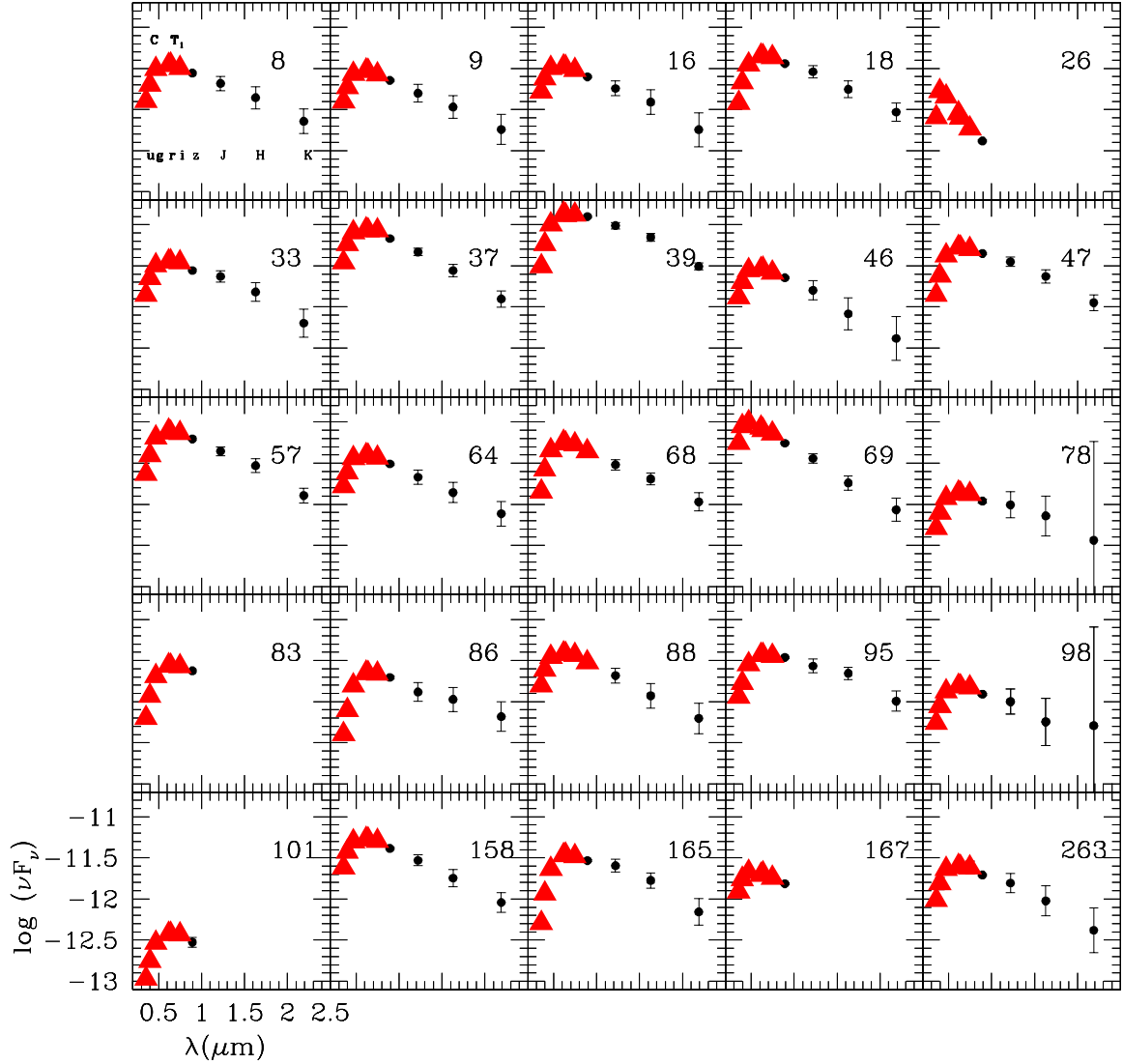} 
\vspace{-1.5in}
\caption{
Spectral energy distributions of 25 bright stars in our FOV, with flags from SDSS DR13 \citep{DR13} that indicate the photometric quality has not been compromised in any manner, and $1\%$ uncertainties. The object-number corresponds to Tables 2 \& 3. Our photometry, scaled to that of SDSS DR13, is shown for the $CT_1ugri$-filters (red filled triangles), where the error bars are smaller than the points. The SDSS-z and 2MASS $JHK$-magnitudes are shown as filled circles. All the plots shown are on the same logarithmic scale, apparent flux density against wavelength in microns.  }
\label{fig:f4}
\end{figure}

\subsection{SDSS Filters}
Along with Washington $(C-T_1)$, the SDSS-color, $(u-g)$, is the most sensitive to age and metallicity for RGB stars, but it is much more sensitive at low ($\mathrm{[Fe/H]}<-2$ metallicities \citep{hug14}. The advantage of using SDSS filters over the Washington colors is having the standards in the field for relative photometry. However, the u-band is centered at 355-nm with a width of only 57-nm, compared with a center of 398-nm and a width of 110-nm for C, making the former require much more observing time. For catalog data from SDSS DR13, the Seg~3 region is complete for $g\approx 23.0\; mag.$ for the $(g-r)$ CMD, but only $g\approx 18.9\; mag.$ for the $(u-g)$-color, when using sources with uncertainties less than 0.25 in
the CMD.

 \begin{figure}
\includegraphics[scale=0.6]{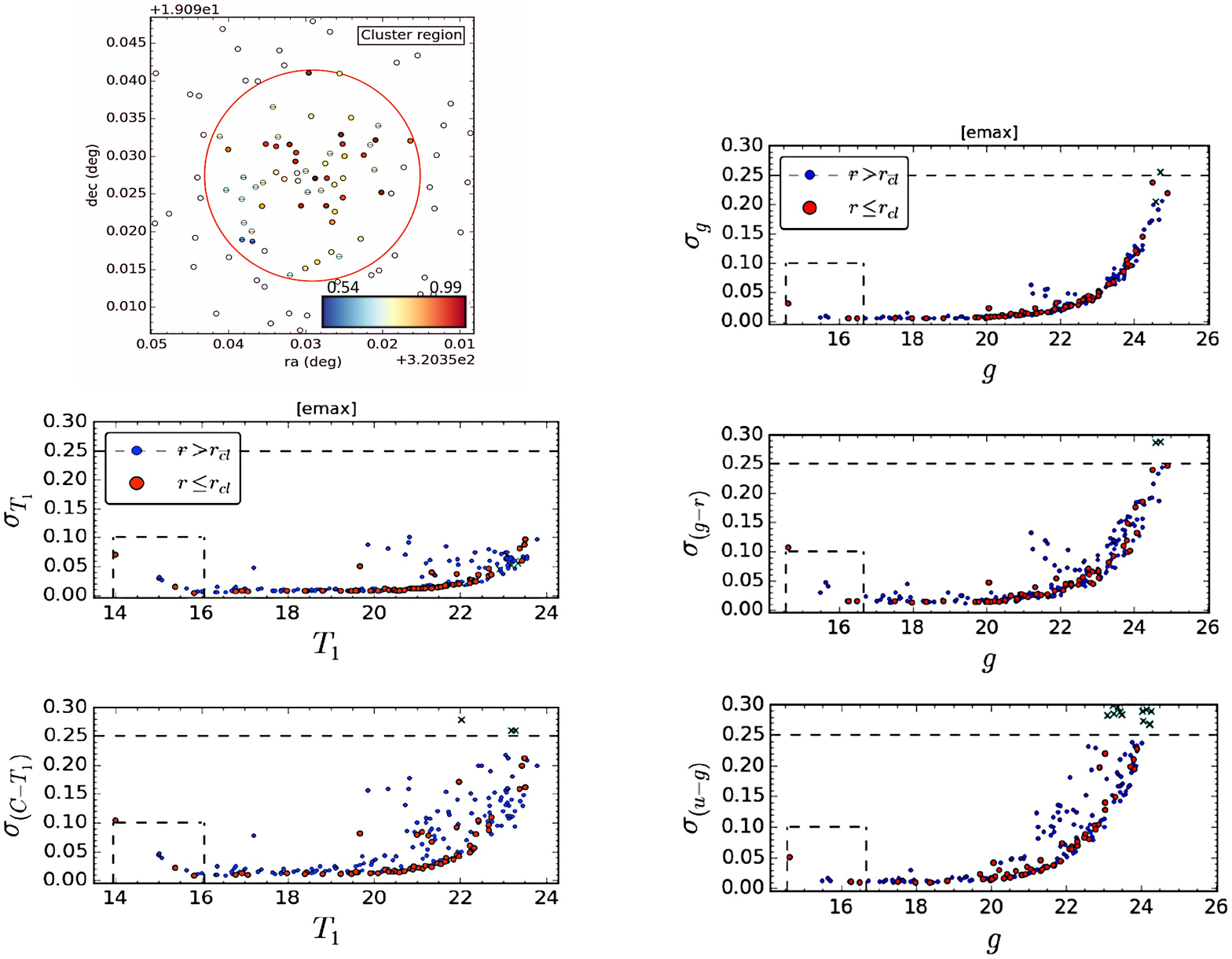} 
\caption{Plots of uncertainty vs. magnitude for the sources detected in the FOV shown in Figure~3, with the cluster radius taken from the O13 data in Figure~1. The cluster ($r_{cl}\sim 0.^\circ 014$) shown in the upper-left panel, with the probability of membership compared to the off-cluster area. The left panels are the magnitudes/colors used for Washington filters and the right panels show the SDSS magnitudes/colors.}
\label{fig:f5}
\end{figure}

\begin{figure}
 \hspace{0.5in}
  \vspace{-1.0in}
\includegraphics[scale=0.75]{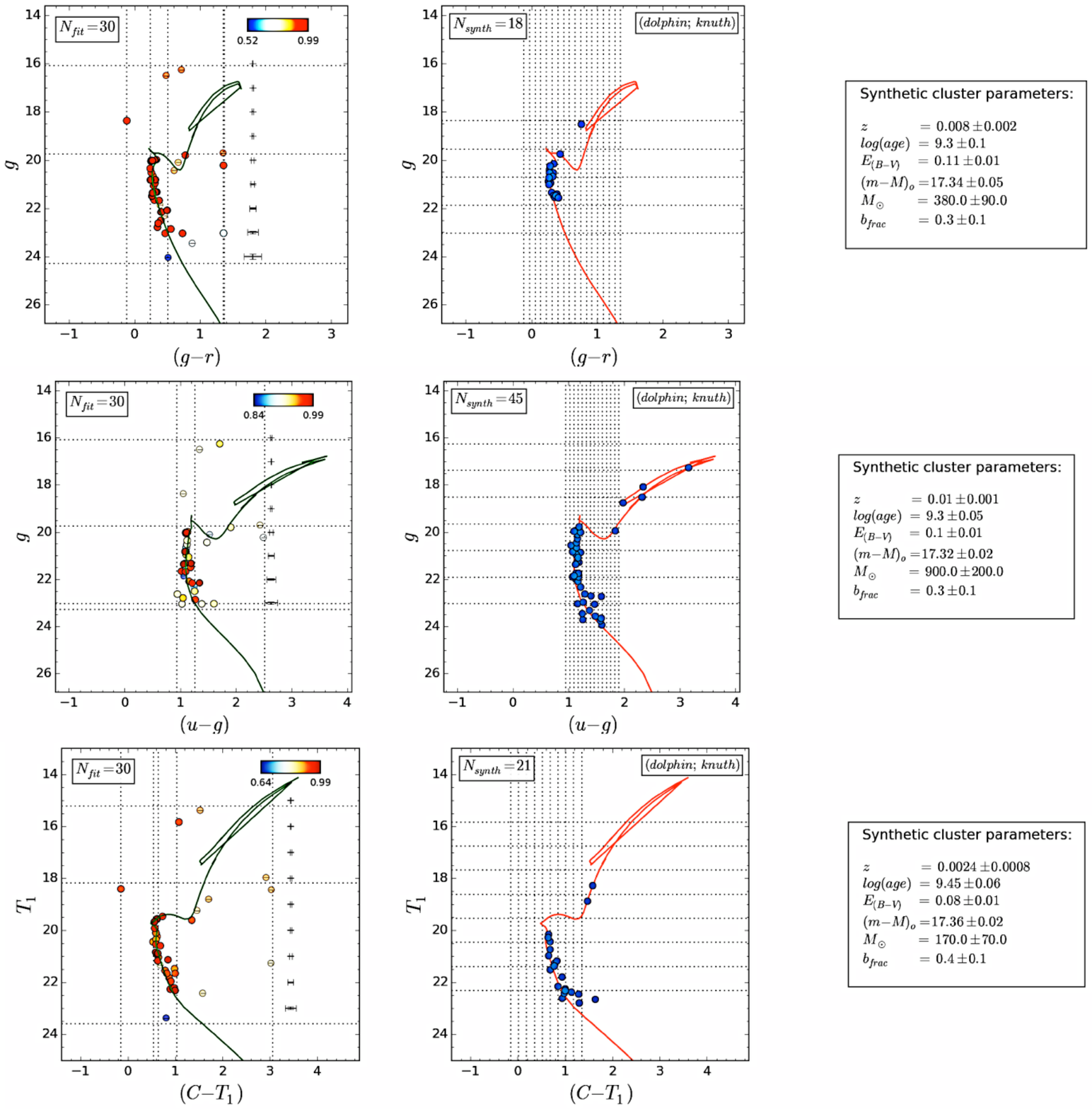}
\caption{We show 3 of the 75 total ASteCA-code runs that fit the Padova isochrones as examples in each color-magnitude diagram, some of which we ran in both manual and automatic modes. The color-coding of each point represents its probability of cluster membership. Where we maximize the number of cluster members and there is the lowest level of background filtering, about $40$ objects are identified as likely members within $r_{cl}\sim 0.^\circ014$. Other cluster members are found outside the cluster radius (F11 \& O13). When we set a medium to high level of background filtering and use the lower estimate of the cluster radius, $\sim 30$ objects are found within $r_{cl}\sim 0.\degr010$, with the tidal radius varying from $0.\degr04$ to $0.\degr1$. These examples show the smaller cluster radius which yields a larger, and more uncertain, tidal radius. Removing the SDSS-identified galaxies does not affect the results significantly as most of them are not in the MSTO region, and are normally rejected by the code.
 }
\label{fig:f6}
\end{figure}

\section{APO Observations}

We used the Apache Point Observatory (APO)'s new ARCTIC Imager and the camera it replaced, SPICam for our observations with the 3.5-m telescope.  The ARCTIC camera has a $4096\times 4096$ STA chip giving $7.\arcmin5\times 7.\arcmin5$ as the FOV, when the new 5-inch diameter circular filters are used. The older Washington filters are $3\arcsec \times 3\arcsec$ and vigniette the FOV. SPIcam had a FOV of $4.\arcmin8 \times 4.\arcmin8$. We have several filter wheels that can handle up to ten $3\times 3$-inch square filters (fewer in full-field mode), where binning $1\times 1$ yields 0.11 arcseconds/pixel.  The fastest readout time in $2\times 2$ binned mode is about 5 seconds. The blue-UV sensitivity of ARCTIC is greater  than that of SPIcam, which was a backside-illuminated SITe TK2048E $2048 \times 2048$ pixel CCD with 24 micron pixels, which we also binned ($2 \times 2$), giving a plate scale of 0.28 arcseconds per pixel. Where we combined the datasets, we binned ARCTIC $2\times 2$ and slightly degraded its resolution. We found no irreducible color-terms between frames taken with both imagers, internally. From 2013 to 2015, we had 11 half-nights total, and 102 frames had seeing better than $2\arcsec$, many of which were under photometric conditions, and several nights had sub-arcsecond seeing. Some of the observations were repeated between SPIcam and ARCTIC, which served to test the new imager.

\begin{deluxetable}{llcrcc}
\tablecaption{Frames Used$^a$}
\tablehead{
  \colhead{UT$^b$} &  \colhead{Imager}& \colhead{Filter$^c$}& \colhead{t(s)} &  \colhead{Airmass$^d$}&  \colhead{FWHM ($\arcsec$)}
  }
\startdata  
2013-08-17 & SPIcam & g & 300.0 & 1.03& 0.9 \\
2013-08-17 & SPIcam & g & 300.0 & 1.03 & 0.8 \\
2013-08-17 & SPIcam & g & 300.0 & 1.06 & 0.6 \\
\enddata
\tablenotetext{a}{Full table online}
\tablenotetext{b}{Year-month-day}
\tablenotetext{c}{R is converted to Washington $T_1$}
\tablenotetext{d}{Effective airmass}
\end{deluxetable}

We observed Seg~3 in $CT_1$ and $ugri$ filters with both SPIcam and ARCTIC. The frames used are listed in Table 1, the overlap between this paper and the $V_r$-data from F11 (not the g- and r-mag. values) and O13 is detailed in Table 2. Our photometry is presented in Table 3 for all 218 objects detected in our field-of-view (FOV) in $CT_1ugri$-filters, where we required detections in all filters in order to produce spectral-energy distributions. We include the z-filter from SDSS DR13 and any 2MASS objects detected, for completeness. We compare the FOV of O13 and the multi-wavelength data we collected in Figure~3a and the source density for our sample in Figure~3b, showing the asymmetry of Seg~3. 

We both reduced and analyzed each night's data separately and then median-filtered the images, weighted according to FWHM of each image, scaled appropriately, to obtain the best source-list for the FOV. The Washington photometry was calibrated to the standard system using the Washington Standard fields \citep{gei96} and absolute photometry. For the SDSS filters, we performed relative photometry with the SDSS-catalog objects in the FOV. Within IRAF we used  \em zerocombine \rm with \em ccdproc \rm to correct the flats and object frames, then \em flatcombine \rm and \em ccdproc \rm to flatfield the object frames. We then rotated, aligned and matched the ARCTIC and SPIcam images (making the former fit the latter's coordinate system) for each filter. We used the \em DAOPHOT \rm program suite within IRAF, using $\sim 30$ stars to create a point-spread function (PSF) for each image, and employed 2 runs of \em allstar \rm to find all sources. We found that this field is not as crowded as a normal GC when artificial star tests were performed, and that ASteCA's completeness limits are consistent with the manual experiments. For the Washington filters, we took the photometry measurements of 3 different nights for each star and averaged them (weighted by errors) to use in transforming magnitudes from the co-added images to the standard system. We constructed a plate solution to match with SDSS-catalog objects, and also compared these with the F11 and O13 measurements. All the results are summarized in Tables 2 \& 3. Comparing our data set with those in the literature, there are 38 stars in both our observations and F11,  176 stars in both our observations and O13, and 38 stars covered by all 3 studies. In Table 2, 27 stars have $V_r$ that indicates they are Seg~3 members, as reported by F11, with $V_r=167\pm 30\; kms^{-1}$. In the SDSS DR13, 176 objects are cataloged, of which 35 are also in the 2MASS survey.

As discussed earlier,  O13 compared their data to F11's photometry, where the published data was dereddened by $E(B-V)=0.1$~mag., and found that F11 magnitudes appear to have a zero-point offset (of $\sim 0.09$~mag.).  We concur with O13's assessment, F11's data is offset to our g- and r-magnitudes also by $\sim 0.1$~mag., where our data was originally calibrated to SDSS Data Release 12, (DR12) but has been updated to DR13 \citep{DR13}.

\begin{deluxetable}{cccccccccc}
\tablehead{\colhead{ID\#}& \colhead{SDSS}& \colhead{$Flag^{*}$}& \colhead{$R.A.$(deg.)}&  \colhead{$Dec.$(deg.)}& \colhead{$X_C^b$}&  \colhead{$Y_C$}&
 \colhead{$V_r(kms^{-1})$}& \colhead{$I_{O13}$}&  \colhead{$V_{O13}$}}
 \startdata
2& J212132.25$+$190527.5& D& 320.384375& 19.090972& 389.819& 14.178& \nodata& 20.167& 21.05\\
3& J212136.46$+$190539.3& G& 320.401958& 19.09425& 176.979& 55.689& \nodata& 20.394& 21.852\\
5& J212131.36$+$190548.8& C& 320.380667& 19.096889& 434.675& 89.991& \nodata& 19.756& 21.372\\
6& J212132.28$+$190552.0& C& 320.3845& 19.097806& 388.314& 101.658& -163.3& 18.806& 19.695\\
\enddata
\tablenotetext{a}{Full table online.}
\tablenotetext{b}{From Figures 3b \& 9.}
\tablenotetext{*}{SDSS DR13 photometric quality grade A--C indicates few warning flags, D--F is untrustworthy, G is a galaxy.}
\end{deluxetable}

Figure~4 is a final sanity check on the photometric calibrations between our SPICam, ARCTIC, SDSS DR13, and 2MASS. We show spectral energy distributions (SEDs) of the 25 brightest stars in our FOV in Figure~3 that were not saturated in SDSS DR13. The $CT_1ugri$-filters (red filled triangles) are displayed, where the error bars are smaller than the points. The SDSS-z and 2MASS $JHK$-magnitudes are shown as filled circles. All the plots shown are on the same logarithmic scale, apparent flux density vs. wavelength in microns.

Since the FOV contains a sparse cluster and (we expect) numerous binaries (O13), we ensured that the stars did not have SDSS-flags from SDSS DR13 warning that the photometric quality was compromised in any manner, and no greater than $1\%$ uncertainties. The object-number corresponds to Tables 2 \& 3.  The $CT_1$-filters, calibrated to the Washington Standard Fields \citep{gei96} did not need any additional offsets (other than the standard conversion) to convert its ``Vegamag" system to flux density, and matched well with the SDSS filters (ABmag). The SDSS-z and 2MASS $JHK$-magnitudes are displayed as filled circles, and have larger error bars.  We first had calibrated linearly to SDSS DR12 and then re-calibrated to DR13 using the photometry for the stars shown in Figure~4, and found that \rm the \rm i-band had a color term.
\begin{eqnarray}
i_{DR13}=i_{DR12}-0.4194*(r-i)_{DR12}+0.1221,\; \sigma_{\bar{x}}=0.010;\\
 r_{DR13}=r_{DR12}+0.0151,\; \sigma_{\bar{x}}=0.003;\\	
g_{DR13}=g_{DR12}+0.0144,\; \sigma_{\bar{x}}=0.003;\\     
u_{DR13}=u_{DR12}-0.004, \;  \sigma_{\bar{x}}= 0.010.\\ 
\end{eqnarray}

\begin{deluxetable}{lrrrrrrrrrrrrrr}
\tablecaption{Segue 3: Sources with Detections in All Filters$^a$}
\tablehead{\colhead{ID\#}&      \colhead{$T_1$}& \colhead{$\sigma_{T_1}$}& \colhead{C}& \colhead{$\sigma_C$}& \colhead{u}& \colhead{$\sigma_u$}& 
\colhead{g}& \colhead{$\sigma_g$}& \colhead{r}& \colhead{$\sigma_r$}& \colhead{i}& \colhead{$\sigma_i$}& \colhead{$z^b$}& \colhead{$\sigma_z$}
   }
\startdata
2& 20.640& 0.010& 21.972& 0.022& 22.683& 0.033& 21.396& 0.015& 20.903& 0.025& 20.668& 0.037& 20.42& 0.14 \\
3& 21.117& 0.031& 23.278& 0.055& 23.666& 0.088& 22.414& 0.046& 21.359& 0.050& 21.292& 0.074& 20.34& 0.16 \\
5& 20.519& 0.009& 23.258& 0.041& 24.377& 0.124& 22.026& 0.017& 20.789& 0.018& 20.685& 0.028& 20.25& 0.12 \\
6& 19.259& 0.007& 20.739& 0.011& 21.588& 0.016& 20.099& 0.006& 19.467& 0.014& 19.229& 0.022& 19.05& 0.05 \\
7& 19.122& 0.011& 20.886& 0.017& 21.859& 0.019& 20.121& 0.008& 19.471& 0.029& 19.225& 0.042& 18.92& 0.04 \\
  \enddata
  \tablenotetext{a}{Full data table online}
  \tablenotetext{b}{Taken from SDSS DR13}
\end{deluxetable}

These uncertainties were added in quadrature with each object's photometric uncertainties from DAOPHOT; the final uncertainties are listed in Table 3. The i-band magnitudes are only used in the SEDs and not for any model-fitting, as $(r-i)$ is a temperature-sensitive color. The range of the $(u-g)$, $(g-r)$, and $(r-i)$ colors for the 25 stars was $+1.05 \; to \; +2.11$, $-0.12 \; to \; +0.82$, and $0.00 \; to \; +0.29$ mag., respectively. We included star $^\#26$ in our calibration since it is the only A-type star in the sample, noting that it is in the cluster-center and the crowded field might affect the photometry in the SDSS catalog (we gave it a C-grade, see Table 2).

The uncertainties versus magnitudes for sources inside (red circles) and outside (blue circles) the cluster radius (from the ASteCA fit for O13) are shown in Figure~5, with objects too faint to be used in any fit shown as cyan crosses. We required that the objects used be detected in all the $CT_1ugri$-filters to allow for SED checks, but we rejected with any object with large photometric uncertainties ($\sigma > 0.25$) in any color in the following analyses.

\subsection{Padova Fits for Washington \& SDSS Colors}
 
In Figure~6, we show 3 of the runs for the combined data set, showing the extremes of the fit to the Padova model grid -- the $(u-g)$ results indicate higher metallicities, which may mean this color is more sensitive to $\alpha$-abundances in general, but also the CMD below the MSTO is almost vertical, and hides the binary sequence that is much clearer in $(V-I)$ and $(C-T_1)$. The 3P King model fit for $(V-I)$ was consistent for the smaller FOV, but the tidal radius was harder to limit without more data away from the center of Seg~3. Our completeness limits are $T_1\approx 23.1$ for the Washington CMD, and $g\approx 21.8$ for the $(g-r)$ CMD (worse than DR13), but this was because we required at least a $5\sigma$ detection in u-band. From 75 runs of the ASteCA code in all colors, iterating toward agreement between the O13, Washington, and SDSS CMDs, the Padova/PARSEC12 solar-scaled models show that Segue 3 has $Z=0.006 \pm 0.002$, $\log(Age)=9.38 \pm 0.11$, $(m-M)_0=17.33 \pm 0.08$, $E(B-V)=0.09\pm 0.01$, and a binary fraction of $0.36\pm 0.12$. The mass estimate from the King models was quite uncertain, $630\pm 264 \; M_\odot$. If we only use  runs where we remove the SDSS-identified galaxies, and use the information from F11 on radial-velocity members: $Z=0.006 \pm 0.001$, $\log(Age)=9.42 \pm 0.08$, $(m-M)_0=17.35 \pm 0.08$, $E(B-V)=0.09\pm 0.01$, and a binary fraction of $0.39\pm 0.05$, with a cluster mass of $478\pm 56 \; M_\odot$ for the synthetic CMD generated with these parameters. With $Z=0.006$ and an age of $2.6^{+0.6}_{-0.4}$ Gyr, our estimates are younger and more metal-rich than O13, using the same set of isochrones. Converting metallicity to $\mathrm{[Fe/H]}\approx -0.5$ if the cluster is not $\alpha$-enhanced, and would be $\mathrm{[Fe/H]}\approx -0.8$, if $[\alpha/Fe]=+0.4$, which we discuss in the next section.

\newpage
 \subsection{Dartmouth Models}
 
  \begin{deluxetable}{cc|ccc|ccc|ccc}
\tablecaption{Dartmouth Models}
\tablehead{\colhead{[Fe/H]} &  \colhead{$[\alpha/Fe]$} &     \colhead{$Y^a$}&  \colhead{$Z^b$}&  \colhead{$Z_0^c$}&  \colhead{Y}& \colhead{Z}& \colhead{$Z_0$}&  \colhead{Y}& \colhead{Z}& \colhead{$Z_0$}}
\startdata
-2.5& 0.0& 0.245&   0.00005& 0.00005&    0.33& 0.00005& 0.00005&    0.4& 0.00004& 0.00004\\
-2.0& 0.0& 0.245&   0.00017& 0.00017&    0.33& 0.00015& 0.00015&    0.4& 0.00014& 0.00014\\
-1.5& 0.0& 0.246&   0.00055& 0.00055&    0.33& 0.00048& 0.00048&    0.4& 0.00043& 0.00043\\
-1.0& 0.0& 0.248&    0.00172& 0.00172&    0.33& 0.00153& 0.00153&    0.4& 0.00137& 0.00137\\
-0.5& 0.0& 0.254&    0.00537& 0.00537&    0.33& 0.00482& 0.00482&    0.4& 0.00431& 0.00431\\ \\
-2.5& 0.4& 0.245&    0.00011& 0.00006&    0.33& 0.00010& 0.00005&    0.4& 0.00009& 0.00004\\
-2.0& 0.4&  0.246&   0.00035& 0.00018&    0.33& 0.00031& 0.00016&    0.4& 0.00028& 0.00014\\
-1.5& 0.4&  0.247&    0.00111& 0.00056&    0.33& 0.00098& 0.00050&    0.4& 0.00088& 0.00045\\
-1.0& 0.4&  0.251&    0.00346& 0.00176&    0.33& 0.00310& 0.00158&    0.4& 0.00278& 0.00141\\
-0.5& 0.4&  0.262&    0.01069& 0.00544&    0.33& 0.00970& 0.00494&    0.4& 0.00869& 0.00400\\
\enddata
\tablecomments{Models from the Dartmouth Stellar Evolution Database (D08).}
\tablenotetext{a}{$Y=0.245+1.6Z$}
\tablenotetext{b}{$Z=$ Z-value of matching isochrone in CMD-photometric metallicity which includes $\alpha$-abundances and Y-effects.}
\tablenotetext{c}{$Z_0=$ expected Z-value, derived iron-abundance from spectroscopy alone.}
\end{deluxetable}

 Table 4 shows a representative sample of the Dartmouth models used, and examples of how they translate to the metallicity scale, with Z, $\alpha$-abundances, and helium or Y-variations.   For comparison, O13 quote a best-fit Padova/PARSEC isochrone of Z=0.003 and an age of 3.2 Gyr. The PARSEC12 models are solar-scaled, but the Dartmouth models (Dotter et al. 2008; hereafter D08) can be solar-scaled or have variable $\alpha$-abundances or Y-content, but only extend to the helium-flash. Recently, \citet[hereafter: D16]{dot16} released solar-scaled models which include all evolutionary stages, which we will refer to as the MIST models (MESA Isochrones \& Stellar Tracks); also see \citet{cho16}. We obtained these models in the Washington and SDSS filter-systems for this project.

 \begin{figure}
 \hspace{1in}
\includegraphics[width=0.7\textwidth]{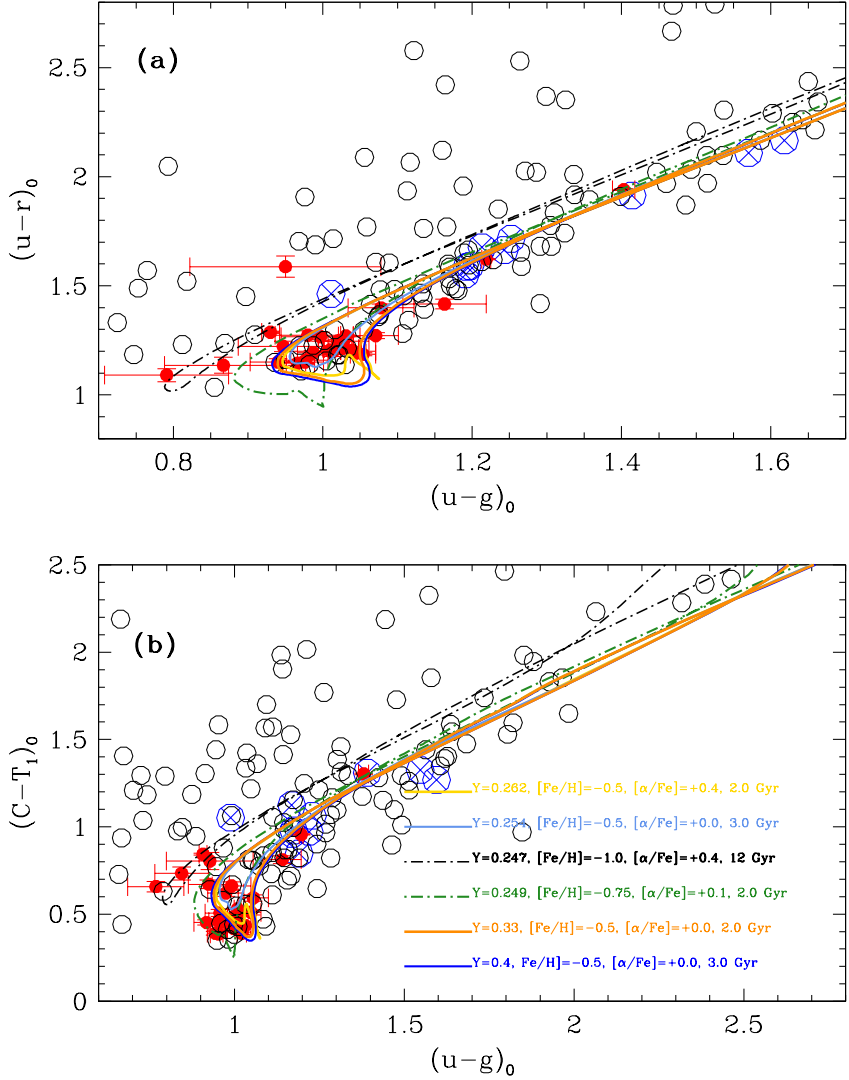}
\caption{For the data from Table 3, we plot the 218 detected objects as black open circles.  Objects identified as radial velocity members by F11 are shown as red filled circles, with error bars. Stars which F11 showed as non-members are shown as blue circles with crosses.  \rm (a) \rm Color-color plot for $(u-r)_0 \; vs. \; (u-g)_0$. \rm (b) \rm $(C-T_1)_0 \; vs. \; (u-g)_0$ All possible Dartmouth models fitting the MSTO are shown for $E(B-V)=0.09$ (O13). }
\label{fig:f7}
\end{figure}

Figure~7 shows color-color plots for our sample, with Figure~7a $(u-r)_0 \; vs. \; (u-g)_0$, and (b) $(C-T_1)_0 \; vs. \; (u-g)_0$. According to the results reported by \citet{li08} on AB-system color pairs:
 $ [(u-r), (r-K_s)]$, $[(u-r), (i-J)]$, and $[(u-K_s), (z-K_s )]$ are more suitable for constraining stellar-population parameters than many others, but we do not have enough good 2MASS detections here, below the RGB. The objects with smaller photometric uncertainties cluster with the more-metal rich MIST models, but the results are not conclusive from the color-color plots alone.

   \begin{figure}
\hspace{+1.0in}
\includegraphics[scale=0.6,clip,trim=0.0in 0in 0.65in 0.3in]{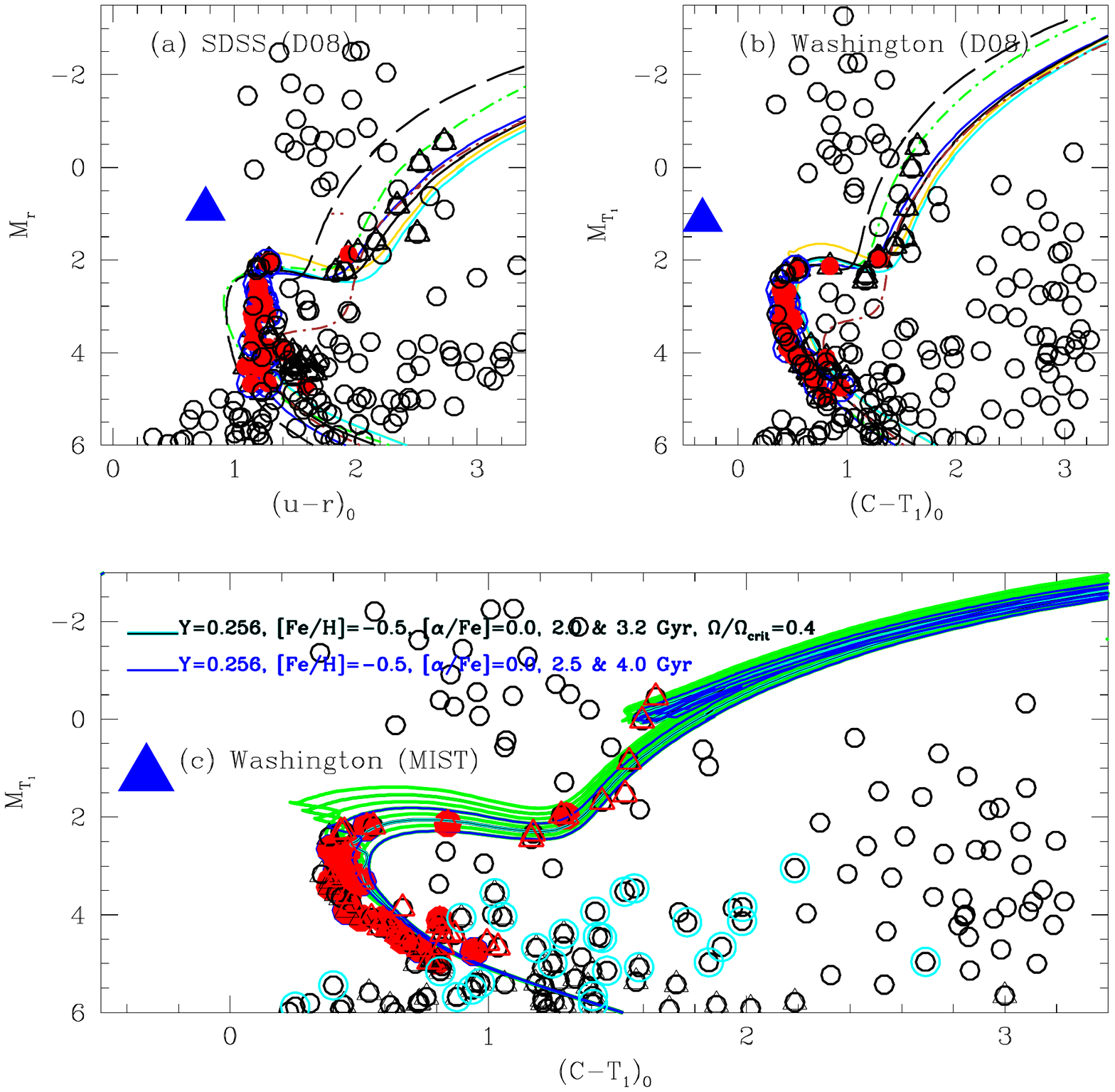}
\caption{For the data in Table 3, we plot the 218 detected objects as black open circles in all diagrams.  Objects identified as radial velocity members by F11 are shown as red filled circles, for which we show the error bars. Stars which F11 showed as non-members are shown as blue circles with crosses. We recovered 19/27 of the F11 members in our sample  when we determined the best-fit D16 isochrone and flagged the objects within $5\sigma $ in color and $7\sigma $ in magnitude of the model. We identified 48 stars (plus a BS) as being associated with the best-fit MIST-models to the F11 completeness limit, about 16 more than the F11 sample and 104 total objects (black open triangles). \rm (a) \rm $M_r \; vs. \; (u-r)_0$ and the D08 models. Object $\# 26$  is a possible blue straggler (blue filled triangle), located in the cluster center, which is never removed by statistical cleaning via ASteCA, so we also included this object. In both (a) and (b), the green line is the model: $ Y=0.249, \; Z=0.002,\; 3.75 \;Gyr, \; \mathrm{[Fe/H]}=-1.0,  [\alpha/Fe]=+0.2$. The blue line is: $ Y=0.4, \; Z=0.004,\; 3.0 \;Gyr, \; \mathrm{[Fe/H]}=-0.5,  [\alpha/Fe]=0.0$.
The cyan line is: $ Y=0.254, \; Z=0.005,\; 3.0 \;Gyr, \; \mathrm{[Fe/H]}=-0.5,  [\alpha/Fe]=0.0$.
The red line is: $ Y=0.33, \; Z=0.005,\; 3.0 \;Gyr, \; \mathrm{[Fe/H]}=-0.5,  [\alpha/Fe]=0.0$.
The brown line is: $ Y=0.25, \; Z=0.003,\; 12.0 \;Gyr, \; \mathrm{[Fe/H]}=-1.0,  [\alpha/Fe]=+0.4$.
The black line is: $ Y=0.249, \; Z=0.002,\; 4.0 \;Gyr, \; \mathrm{[Fe/H]}=-1.5,  [\alpha/Fe]=+0.8$.
The yellow line is: $ Y=0.257 \; Z=0.007,\; 2.5 \;Gyr, \; \mathrm{[Fe/H]}=-0.5,  [\alpha/Fe]=+0.2$.
\rm (b) \rm $M_{T_1} \; vs. \; (C-T_1)_0$ and the D08 models.
\rm (c) \rm $M_{T_1} \; vs. \; (C-T_1)_0$ and the D16/MIST models. The 28 objects that are identified as galaxies in SDSS DR13 are circled in cyan. A  $\sim$single-metallicity Dartmouth model-grid is shown, fitting the MSTO, for $(m-M)_0=17.32$ and $E(B-V)=0.09$ (O13): $ Y=0.246, \; Z=0.005, 2-4\; Gyr, \; \mathrm{[Fe/H]}=-0.5,  [\alpha/Fe]=0.0$. In addition, the objects within $5\sigma $ in color and $7\sigma $ in magnitude of the model, above the limit for the F11 sample (red filled circles) are shown as red open triangles, and those below that limit are shown as black open triangles. Object 26 is identified as a blue filled triangle. SDSS-galaxies are shown as cyan open circles. }
\label{fig:f8}
\end{figure}

In Figure~8a, the $M_r \; vs.\; (u -r)_0$ CMD with the D08 models, object $^\#26$ is the likely BS  (blue filled triangle), located in the cluster center.  The objects within $5\sigma$ in color and $7\sigma$ in magnitude of the average of the best-fit models ($ Y=0.246, \; Z=0.005, 2-4\; Gyr, \; \mathrm{[Fe/H]}=-0.5,  [\alpha/Fe]=0.0$), above the limit for the F11 sample (red filled circles) are shown as red open triangles (49 objects), and those below that limit are shown as black open triangles (an extra 54 sources).  SDSS-galaxies are shown as cyan open circles. There are a total of 103 sources enclosed by the 2.0--3.2~Gyr isochrones. When we take those 49 stars from the restricted CMD, and re-sample the models,  setting the distance and extinction: $Z=0.004\pm 0.002$, which translates to $\mathrm{[Fe/H]}=-0.65^{+0.05}_{-0.09}$, with an age range of $\log (Age)=9.35^{+0.10}_{-0.35}$ (from quartiles around the median) but the mean is $\log (Age)=9.05\pm 1.09$.  The most representative statistic of the MSTO stars is the median $\pm$ the $1^{st}$- and $3^{rd}$-quartile values:  $2.2^{+0.6}_{-1.2}$ Gyr -- younger than the D08 values.
 We can recover 19/27 of the F11 members in our MIST-sample by this selection method (Figure 8). Several of the upper-MS F11 stars were not recovered because they were more than $5\sigma $ away from the fiducial line (cyan/black) in $(C-T_1)$ and their uncertainties were small -- which may be an indication of above-average stellar rotation (see \S 5).  This method under-counts the membership below the MSTO, but excludes the RGB stars which are $V_r$ non-members (F11), and galaxies (cyan en circles) -- giving us confidence that most of the new objects (identified from MIST-matching) above the MSTO  belong to Segue 3, and we will target these with future spectroscopy.

Figure~9 is the final source-map for our Seg~3 data. All the 103 objects selected as MIST-matches are shown as black open triangles, with the objects above the F11-limit over-plotted as heavy grey open triangles. All stars detected in the FOV are shown as black open circles, scaled by $T_1$-magnitudes; the 218 detected objects from Table 3 are plotted as  open black circles scaled by $T_1$-mag. The 28 SDSS DR13 galaxies are circled in cyan.  F11's confirmed members are shown as red filled circles and non-members are shown as blue crosses, on top of the red filled circles. Object 26, the BS-candidate, is shown as a blue filled triangle. The cluster radius is shown as a red circle (as in Figure~1); the average tidal radius is shown as a dashed red circle. Grey-blue circles show successive $30\arcsec$ radii from the center of the cluster. 
\begin{figure}
\hspace{+1.0in}
\includegraphics[width=0.7\textwidth, height=0.8\textwidth]{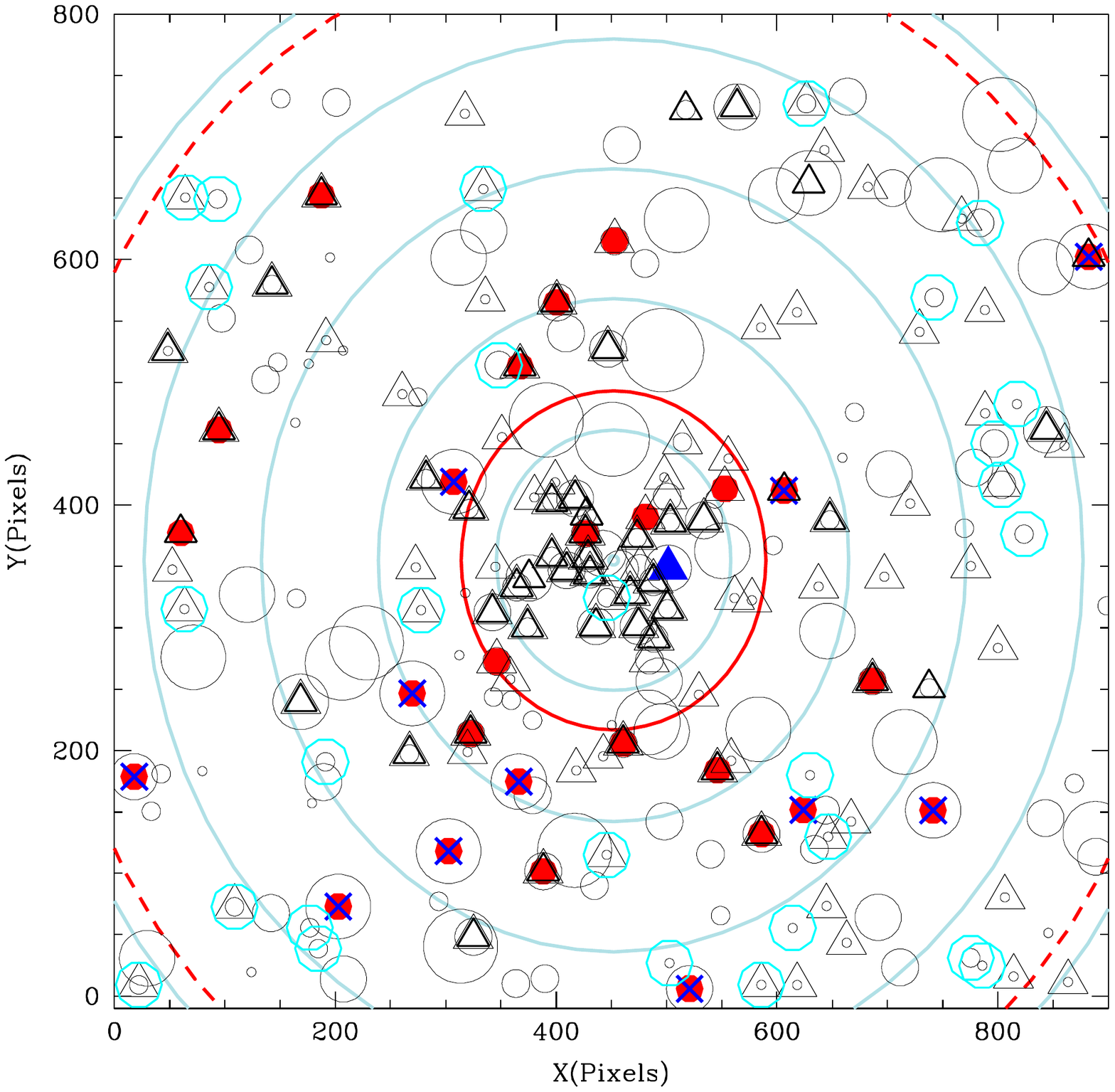}
\vspace{-1.0in}
\caption{The 218 detected objects from Table 3 are plotted as open black circles, scaled by $T_1$-magnitude. The 28 objects that are identified as galaxies in SDSS DR13 are circled in cyan. Objects that have radial velocities measured by F11 are shown as red filled circles. Stars which F11 showed as non-members are shown as blue crosses. The members identified as being within the MIST isochrones are shown as heavy grey open triangles (within $5\sigma $ in color and $7\sigma $ in magnitude of the models, above the limit for the F11 sample), and those below that limit are shown as fainter open triangles. Object 26 is identified as a blue triangle.  The cluster radius is shown as a red circle; the tidal radius is shown as a dashed red circle. Grey-blue circles show $30\arcsec$ radii from the center of Segue 3. }
\label{fig:f9}
\end{figure}

\section{Spectral Energy Distributions}

We selected the 14 brightest stars above the MSTO from the MIST sample and plot the spectral energy distributions (SEDs) with standard ATLAS9 model fluxes (Castelli \& Kurucz 2003; 2004) in Figure~10. The photometry is taken from Tables 2 \& 3, using V \& I from O13, and \em z \rm  from SDSS DR13. The temperature and surface gravity are estimated from an average of the D08 and D16 models, allowing for either $\mathrm{[Fe/H]}=-0.5$ or $\mathrm{[Fe/H]}=-1.5$. All the red models are for $\mathrm{[Fe/H]}=-0.5$ and $[\alpha/Fe]=0.0$. The blue models are for $\mathrm{[Fe/H]}=-1.5$ and $[\alpha/Fe]=+0.4$, scaled to the same peak flux density. Stars confirmed as $V_r$-members have an asterisk after their identification numbers. Figure~10a shows $0.2<\lambda < 1.0 \; \mu m $ and Figure~10b displays $0.3<\lambda < 0.5 \; \mu m $. Star $^\#26$, the suspected BS, is too hot for the low-resolution ATLAS9 models to show much of a difference in the metallicity-sensitive u- and C-bands. The stars between the MSTO and the base of the RGB, $^\#205^*, \; 143^*,\;  100,\; \& \; 71^*$ are also too hot to show much of a difference in the UV-bands. However, all stars brighter than $6^*$ on the RGB/AGB appear to favor the more metal-rich models. Object $^\#273$ is the only source that shows variability between O13, SDSS DR13, and our photometry, and it is located near the (metal-rich) red horizontal-branch.  All O13 V- \& I-photometry points are surrounded by a blue square, and the z-mag. is from DR13.

\begin{figure}
\hspace{-0.5in}
\includegraphics[scale=0.7]{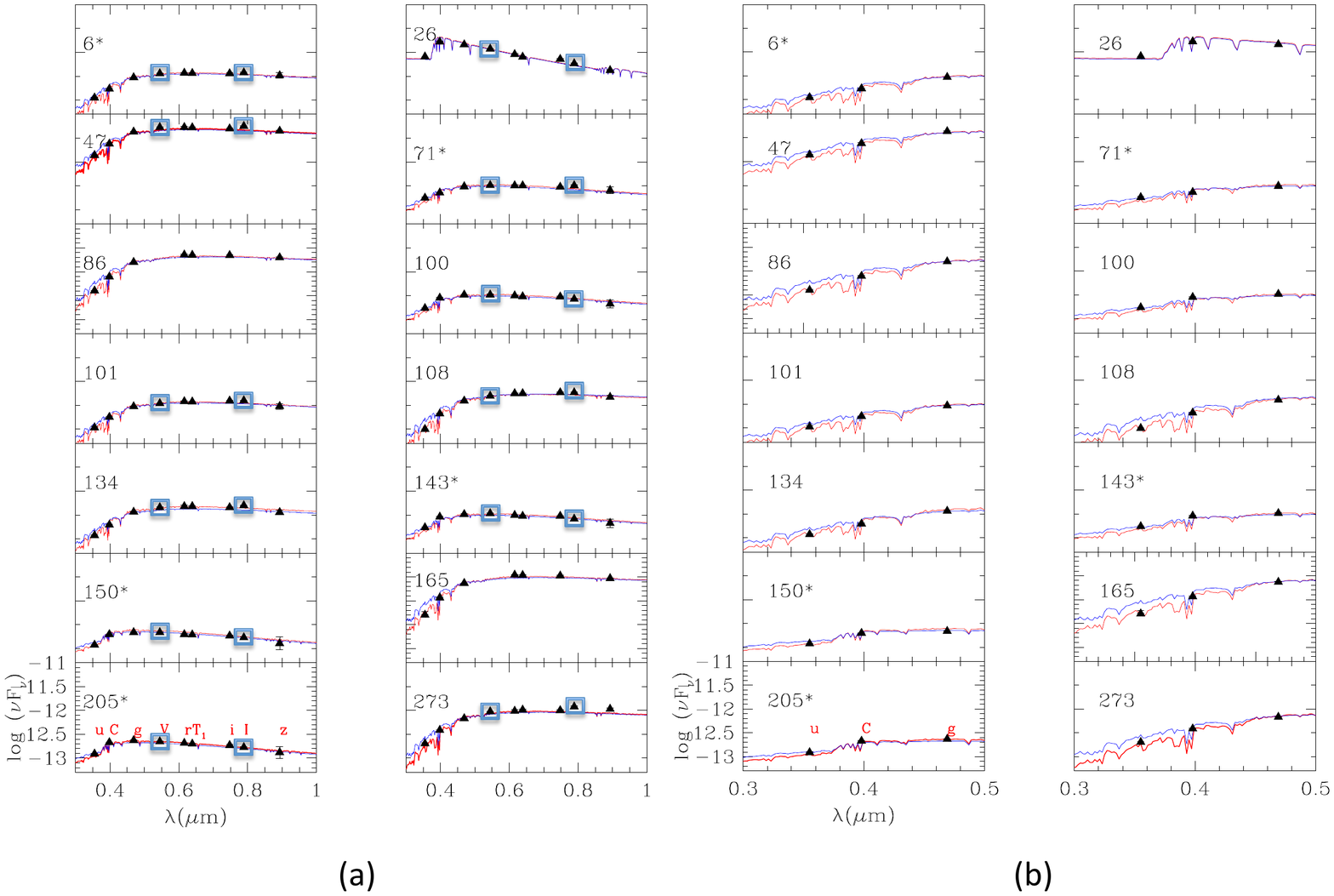}
\caption{Spectral energy distributions for the 14 brightest (likely) Seg~3 members. The magnitudes were converted to fluxes using the normal {\tiny ABMAG/VEGAMAG} \rm calibrations (see Hughes et al. 2014), but no additional zero-point shifts were applied. All photometry is shown as filled black triangles; the V- and I-band  data is (with blue squares) from O13 Table 3's photometry, and the z-filter flux is from SDSS DR13. The data-points have $1\sigma$-error bars (not given in O13), but they are often covered by the symbols.  The source names correspond to the stars listed in Tables 2 \& 3, with the best fit ATLAS9 models shown for $\mathrm{[Fe/H]}=-0.5$ (red: solar-scaled) and $\mathrm{[Fe/H]}=-1.5$ (blue: $\alpha$-enhanced). The y-axes are flux density in $ergs/s/cm$, and the x-axes are wavelength in microns. Objects with asterisks after the number have been confirmed as members by radial velocities. \rm (a) \rm $0.2<\lambda < 1.0 \; \mu m $.
\rm (b) \rm $0.3<\lambda < 0.5 \; \mu m $.
}
\label{fig:f10}
\end{figure}

The three most luminous possible members, $^\#47$, 165, and 86, are separated from the rest of the RGB/SGB MIST-sample, and could be AGB/HB stars. With $\sim 40\%$ binaries, this cluster could very well have evolved massive BS, but an argument against this interpretation is that they are not in the core of the cluster. Their SEDs do appear to be better-fit by the metal-rich models, but these stars require radial-velocity measurements and are bright enough for higher resolution spectral analysis if they are confirmed as members of Seg~3.

\newpage

\section{Age Spread vs. Rotation Range}

\rm The statistical uncertainties in the photometry at the MSTO should not be large enough to mimic an age spread of $>0.5$~Gyr.  O13 compared the Padova and Yale \citep{dem04} isochrones and concluded that their final uncertainty on the age of Seg~3 was  around $\pm 0.5$~Gyr with a maximum age of 4--4.5~Gyr in the $(V-I)$ data, with a mean of 3.2~Gyr.
From the Padova models, with multiple colors to consider, internal and systematic uncertainties, the best fits we can achieve only limit the MSTO age to $2.6^{+0.6}_{-0.4}$ Gyr at $Z=0.006\pm 0.002$. For the 49 objects in the MIST-selection, the quartile-analysis gives  $2.2^{+0.6}_{-1.2}$ Gyr for $Z=0.004\pm 0.001$, where both the MIST and Padova models are solar-scaled. For the D08 models, with $\alpha$-abundances and helium content allowed to vary, the [Fe/H]-grid was coarser: the metallicity-distribution peaks at $Z=0.003\pm 0.002$, $\mathrm{[Fe/H]}=-0.8 \pm 0.4$, with the median being $-0.99$ and the mode=$-0.5$. For the same 49 stars, $[\alpha/Fe]=+0.14 \pm 0.15$, so we can justify using the solar-scaled models. Interestingly, $Y=0.30\pm 0.06$ from D08.
\rm

\newpage
\subsection{\rm Helium Abundances \& Stellar Ages with BASE-9}

\rm We turned to a different Bayesian-analysis code, to reduce systematic uncertainties and handle possible helium-abundance variations.    BASE-9 is a Bayesian modeling code which fits cluster parameters for GCs which differ in helium abundance \citep{ste16}, using the Y-enhanced Dartmouth isochrones (D08). The BASE-9 code requires prior removal of the field star population.

We would expect that, for a constant [Fe/H], increasing the $\alpha$-abundance makes the evolutionary tracks appear fainter and redder, compared to solar-scaled models (increasing Z). If a GC population exhibits differences in Y and other
light elements (CNO), the effect should be detectable in UV CMDs \citep{wag16b}. Among sources of the Y-variation in GCs, \citet{wag16b}  discuss AGB stars, fast-rotating massive stars (FRMS), or pre-main sequence disk-pollution \citep{dec07,dan15}, all of which might be expected to operate in a GC-environment.
In contrast,  \citet{van12} found that the strongest effect on the isochrones was produced by Mg and Si. We can make a direct comparison by requiring the total abundance of heavy elements to be the same value. For an individual star, increasing Y would make the main-sequence lifetime shorter because the rate of H-burning increases \citep{van12}. 

\citet{wag16b} investigated 30 GCs from archival HST data in F275W, F336W, and F438W. Of their sample, 3 clusters have $\mathrm{[Fe/H]}\approx -0.7\; \mathrm{to} \, -0.8$: NGC6624, NGC6637, and NGC6838, with $\log (Age)=9.96,\; 9.98,$ and 10.01. 
The first 2 clusters have $DM\approx 15.3$ and NGC6838 has $DM=13.4$.
Fixing the [Fe/H]-value \citep{wag16b}, 2 populations differing in Y-values were found for each system:
NGC6624 has $Y_A=0.265^{+0.001}_{-0.002}$ and $Y_B=0.343\pm {0.002}$; NGC6637 has  $Y_A=0.265\pm 0.001$ and $Y_B=0.330\pm 0.000$; NGC6838 has $Y_A=0.301\pm {0.003}$ and $Y_B=0.341^{+0.004}_{-0.002}$. The proportion
of Population A stars (lower Y-value) is $\sim {2\over 3}$ for NGC6624 \& NGC6637, and $\sim 0.4$ for NGC6838.  The more metal-rich clusters in \citet{wag16b}'s sample tend have a smaller range in Y-values, and in these HST filters, the RGB-positions are shown to be displaced by increasing $\Delta Y$ in their CMDs. In general, the inner-MWG GCs
 tend to have a higher proportion of Population A stars than the outer-MW, over the whole metallicity range of $-2.37<\mathrm{[Fe/H]}<-0.70$ in this sample.

For the Segue 3 data set, we prepared input files for the BASE-9 program \citep{hip14, ste16,wag16a}, using the statistical cleaning provided by ASteCA, the F11 radial velocities, and the MIST-selection.
We ran BASE-9 for the SDSS and Washington colors and found no strong evidence of a multiple population differing in Y-abundance, with the code not settling down and giving 2 distinct values.  When we did not restrict the input parameters' ranges in any way, the SDSS-colors reasonably matched the Padova models output from ASteCA. The $90\%$ confidence levels gave a range of $9.42   < \log (Age) < 9.57$, $-0.64  <\mathrm{[Fe/H]}<  -0.45$, $17.14  < (m-M)_0 <  17.24$, $0.32 < A_V<  0.35$, and $0.25  < Y <   0.29$. However, the Washington colors gave: $9.04 < \log (Age) <   9.10$,  $-0.54   <\mathrm{[Fe/H]}<  -0.47$,  $18.31  < (m-M)_0 <  18.57$,  $0.29 < A_V<   0.32$, and $0.25  < Y <    0.28$. The Washington colors are more sensitive to metallicity, but there is also a stronger degeneracy effect between age, distance and visual extinction, so the extinction had to be set manually and not allowed to vary. However, if we set the distance modulus and $E(B-V)$ to the O13 values the SDSS colors show an interesting effect in BASE-9: there is no abnormal double-Y population, but there seems to be a double-peak in age, with $-0.89 <\mathrm{[Fe/H]}< -0.69$, as shown in Figure~11. The SDSS colors yield $Y=0.27\pm 0.02$, but imply an age spread of $\sim 0.4$~Gyr, where the 2 peaks are clearly resolved in Figure~11 and the $\log (Age)$ resolution is better than 0.1~dex.

\rm
 
\begin{figure}
\plotone{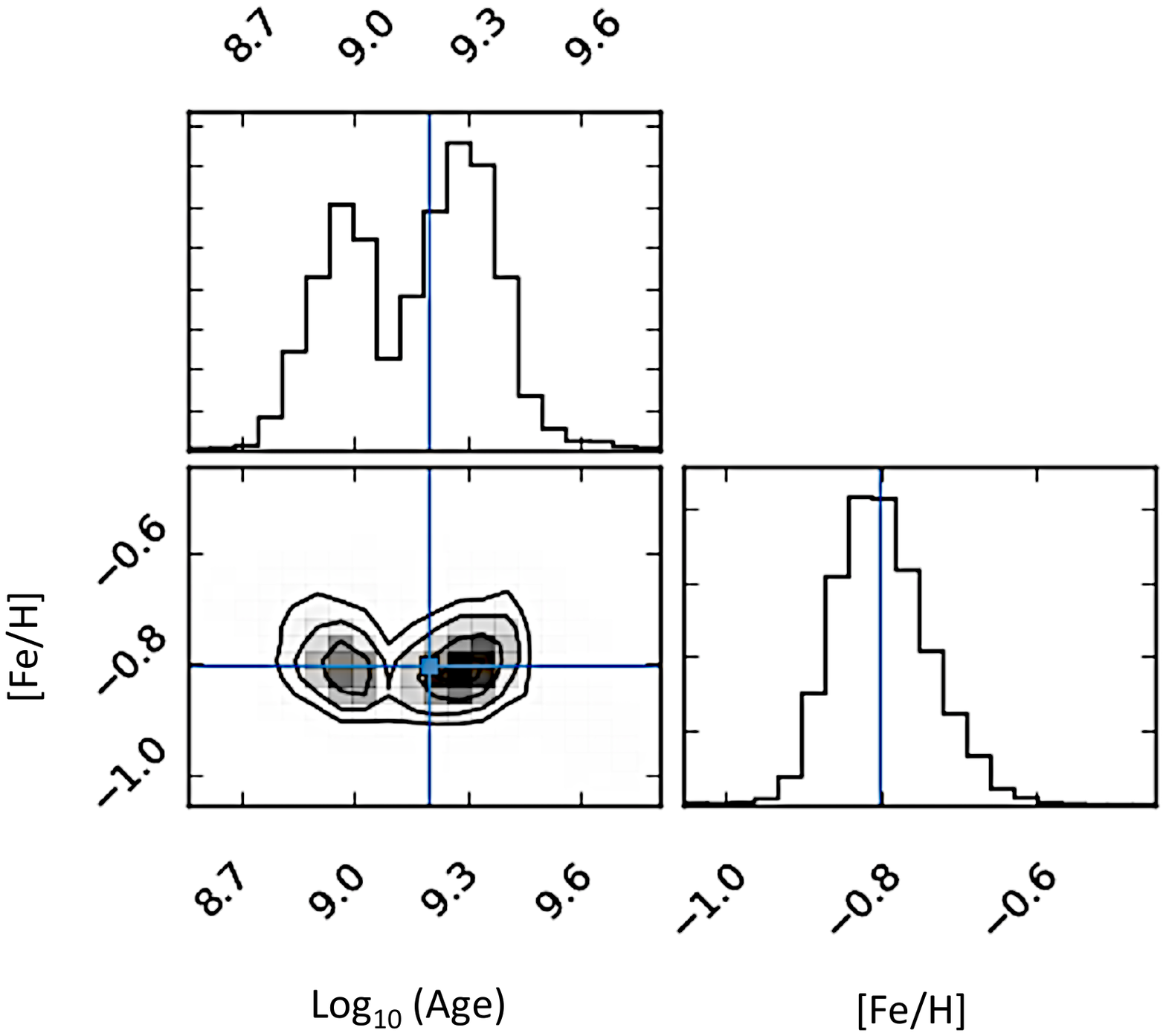}
\caption{BASE-9 results for Seg~3's $(u-g)$ colors. Here, we set $(m-M)_0=17.33$, $E(B-V)=0.09$, where $-0.89 <\mathrm{[Fe/H]}<    -0.69$, Y is not fixed and does not show 2 strong populations, but the age shows 2 peaks. }
\label{fig:f11}
\end{figure}

\subsection{\rm Rotation at the Turn-Off}

In both the Padova and Dartmouth models, with ``chi-by-eye" and Bayesian statistics,  we cannot force the age-distribution uncertainty at the MSTO much below a range of $\pm 0.5$~Gyr.   The narrow SGB would normally preclude any age spread and should argue against any multiple population in Seg~3, characteristic of the ancient MW GC population \citep{gra12}.  Certainly, Seg~3 shows the effect of a large proportion of binaries in the core ($\approx 40\%$), but the luminosity spread at the MSTO could also be the effects of rotation \citep{li12,pia16}.

The closest comparison cluster to Seg~3 in age and metallicity and the shape of the MSTO (admittedly, a much more massive system) can be found in the SMC. \citet{li14} published a Nature article on NGC 1651, an intermediate-age, massive cluster. NGC 1651 had exhibited a supposed age-spread at the MSTO, but \citet{li14} has explained the extended MSTO by stellar rotation variations. Later, \citet{li15} updated their work and found that the best-fit to the observed CMD involved a
period of extended star formation which resulted in population ages from 1.4--1.8 Gyr, that comprised 50\% binaries and 70\% stars with enhanced rotation.

In Figure~12a,  we show the $V_r$ members and the MIST-selected sample's Hertzsprung-Russell diagram using the D08 and D16 models (model fits listed in Table 5), where we number the stars from Figure~10. The cyan/black track is the best fit to the MSTO and the SGB is best fit by the red track with $Y=0.254, \;  \mathrm{[Fe/H]}=-0.5, \; [\alpha/Fe]=0.0, \; 2.5\; Gyr, \; \Omega / \Omega_{crit}=0.4$ (all the MIST models are set to that average rotation here).  The F11-comfirmed members at the MSTO have a mass of $\approx 1.3M_\odot$, in agreement with O13; the unconfirmed post-SGB objects could be more massive only if they are evolved BSs, not if they are normal post-MS stars. The Dartmouth models cover Seg~3's (confirmed) mass range from the upper-MS to the SGB is $M=0.7-1.5M_\odot$ (with a BS estimated mass of $\sim 2.7M_\odot$). Figure~12b is adapted from \citet{li14,geo13}, showing Geneva models for a cluster of the same approximate age and metallicity (NGC 1651 \& Seg~3) with the full range of rotation rates, mass tracks for $M=1.7M_\odot$ are shown. Figure~12c shows a full range of models for a $2.0 M_\odot$ star. In Figure~12b, just at the ``jump" between the MSTO and SGB, the difference of 0.04 dex in $\log T_{eff}$ translates to about 0.5 Gyr in an apparent age difference for the stars, but that can also indicate the difference between a non-rotating star and a turn-off star which is rotating almost at breakup.  This is the clearest result of a spread in rotation but not age, causing the somewhat braided appearance of the MSTO in $(C-T_1)$ (Figure~8c). We conclude that the MSTO morphology is likely to be a combination of rotational effects and the high fraction of binaries because the SGB is too narrow to allow for multiple populations, at least amongst the stars remaining in the core of Seg~3.
The YMCs studied by \citet{li16} in the LMC, NGC 1831, NGC 1868 and NGC 2249, appear to have multiple populations and could show age spreads and broader SGBs, but are all younger than 1.5 Gyr.
\rm Examining the MSTO in Figure~8c compared to Figure~12a, the isochrone-fit has rejected what would have been the slow rotators  -- as we can see from the models in Figures~12b and 12c. Since we recovered 19/27 F11 objects, and most of those non-recovered were bluer than the MIST  $\Omega /\Omega_{crit}=0.4$ isochrone, we surmise that $\sim 70\%$ of the Seg~3 stars have enhanced rotation (leaving $30\%$ with little or no rotation), in a cluster with $\sim 40\%$ of the stars being binaries, which is similar to NGC 1651\citep{li15}. \rm

The Geneva online database can calculate a coarse grid of models for 2 rotation rates (zero and $\Omega /\Omega_{crit}=0.568$) for $Z=0.002$, which would correspond to $\mathrm{[Fe/H]}=-0.9$ and $[\alpha/Fe]=+0.4$ on the D08 scale. When we examined the MSTO masses of the $V_r$-members against the Geneva models, the non-rotating case gives the TO mass as $1.2M_\odot$ and the limit of the stars with radial velocities as about $0.7M_\odot$. The rotating models give these limits as $1.25M_\odot$ and $0.8M_\odot$, which is not significantly different from the D08/D16 models. However, the BS mass is higher than in the D08/D16 system. The Geneva models for $Z=0.002$ can fit the Seg~3 MSTO with an age difference from $\log (Age)=9.45$--9.5, which is 2.8--3.2 Gyr, or  $\log (Age)=9.45$ with a rotation range of $\Omega /\Omega_{crit}=$0--0.95, but the narrow SGB luminosity implies a younger age for the system. The D08, D16, Padova, and Geneva models generally fit Seg~3's CMD for $Z=0.004\pm 0.002$ and ages of $2-3$~Gyr, but the ASteCA code gives us confidence in the uncertainties, rather than a fit by-eye alone.

To further test the rotation hypothesis on Seg~3, we created model clusters with ``The Geneva SYnthetic CLuster Isochrones \& Stellar Tracks" (SYCLIST) code, which can produce synthetic clusters at Z=0.014 and Z=0.002 \citep{geo14,geo13}. We selected the $Z=0.002$ case and produced models shown in Figure~12d \& e, for the appropriate ages.

\citet{geo14} illustrates that for clusters with a metallicity $Z=0.002-0.014$ and an age range of $0.030-1$~Gyr, the SYCLIST models show that the largest proportion of fast rotators on the MS exist just below the turn-off. This happens because rotation extends the  MS lifetime compared to non-rotating models, whereas helium enhancement reduces it,  and the fraction of fast rotators one magnitude below the turnoff also increases with the age of the cluster between 30 Myr and 1 Gyr also. Our best estimates  for the $(C-T_1)$ CMD (see Figure~8c) fit with the D16/MIST models, give $Y=0.256, \;  \mathrm{[Fe/H]}=-0.5, \; [\alpha/Fe]=0.0, \; 2.0-3.2\; Gyr, \; \Omega / \Omega_{crit}=0.4$. For the stars that are about 1 mag. below the MSTO, the stars to the far-red side of the MS are a clear binary sequence and the stars to the far blue could be faster rotators instead of $Y\sim 0.3$ (D08 models). This effect is most obvious in the Washington system. \citet{bra15} also discuss MESA (which we call D16/MIST) rotating models in relation to CMDs for LMC clusters, showing examples of the broad MSTOs reaching a maximum  extent for cluster-ages  of 1--1.5 ~Gyr and tailing-off by $\sim 2$~Gyr, where the models show this in Figure~12c \citep{geo13} for $2M_\odot$ stars. The shape of the Seg~3 MSTO/SGB in Washington colors compared to the MIST isochrones and SYCLIST models puts a lower limit on the age of the Seg~3 cluster at 2~Gyr, but implies a range of rotation rates. The models generated in Figures 12d \& 12e for $\log (Age)= 9.5$ \& 9.45 (light blue and gold, respectively) indicate that the Seg~3 SGB  stars should be younger than 3.0 Gyrs. 
 
 \begin{deluxetable}{ccllllll}
\tablecaption{Input for H-R Diagrams for the Photometric Sample Above the MSTO}
\tablehead{\colhead{ID} &  \colhead{$\log (Age)$} &    \colhead{$Z$}&  \colhead{$\log T_{eff}$}&  \colhead{$\log L/L_\odot$}&  \colhead{$\log g$}&  \colhead{$M/M_\odot$} & \colhead{Comments}  }  
\startdata
   6$^a$ &		 6.00 &	 0.0045& 		     3.718& 	1.113& 	3.464& 	2.043& Brightest SGB\\
   26&	 8.25$^b$ & 	0.0045&		4.095& 	1.887& 	4.304& 	2.609& Possible BS\\
   47 & 	8.70$^b$& 	0.0046& 	    3.720& 	2.271 &	2.419&	 2.623& Brightest Post-MS\\
   71$^a$ & 	6.40$^b$& 	0.0045&   	    	 3.776&		1.055& 	3.727&	1.934& SGB\\
   86 &	 9.35&	 0.0046&    		   3.694&	 1.912& 	2.415&	1.474& Post-MS\\
   100 & 	9.40& 	0.0042&        	3.833& 	1.058& 	 3.792& 	1.335& SGB \\
   101  &	 9.25& 	0.0045&      	3.712& 1.130 & 3.324& 1.522& RGB/SGB \\
   108 & 	5.15$^b$& 	0.0045&     	  3.698&   1.311& 3.093& 1.652& RGB \\
   134 &    9.45& 0.0045&      3.706& 1.241& 3.100& 1.3301& SGB\\
143$^a$  &    9.40& 0.0042&     3.837& 1.052& 3.812& 1.330& SGB\\
  150$^a$ & 	9.35& 	0.0041&        	3.852& 0.868& 4.054& 1.322& SGB\\
  165 &	 9.00&	 0.0046&          3.689& 2.111& 2.354& 2.073& Post-MS\\
    205$^a$ &	 9.30&	 0.0042&        	3.8656 & 0.889& 4.097& 1.349& MSTO\\
  273 &	 9.20&	 0.0046&     3.699 & 1.577& 2.818& 1.603& Post-MS -- Variable?\\
 \enddata
 \tablecomments{Models D16/MIST. We restricted the distance modulus and extinction from our Padova fits and let $Z=0.006\pm 0.002$, with no restriction on age.}
 \tablenotetext{a}{Identified by $V_r$ in F11}
 \tablenotetext{b}{Young age estimate. BS or evolved BS?}
\end{deluxetable}

\begin{figure}
\hspace{-0.4in}
\includegraphics[width=0.6\textwidth]{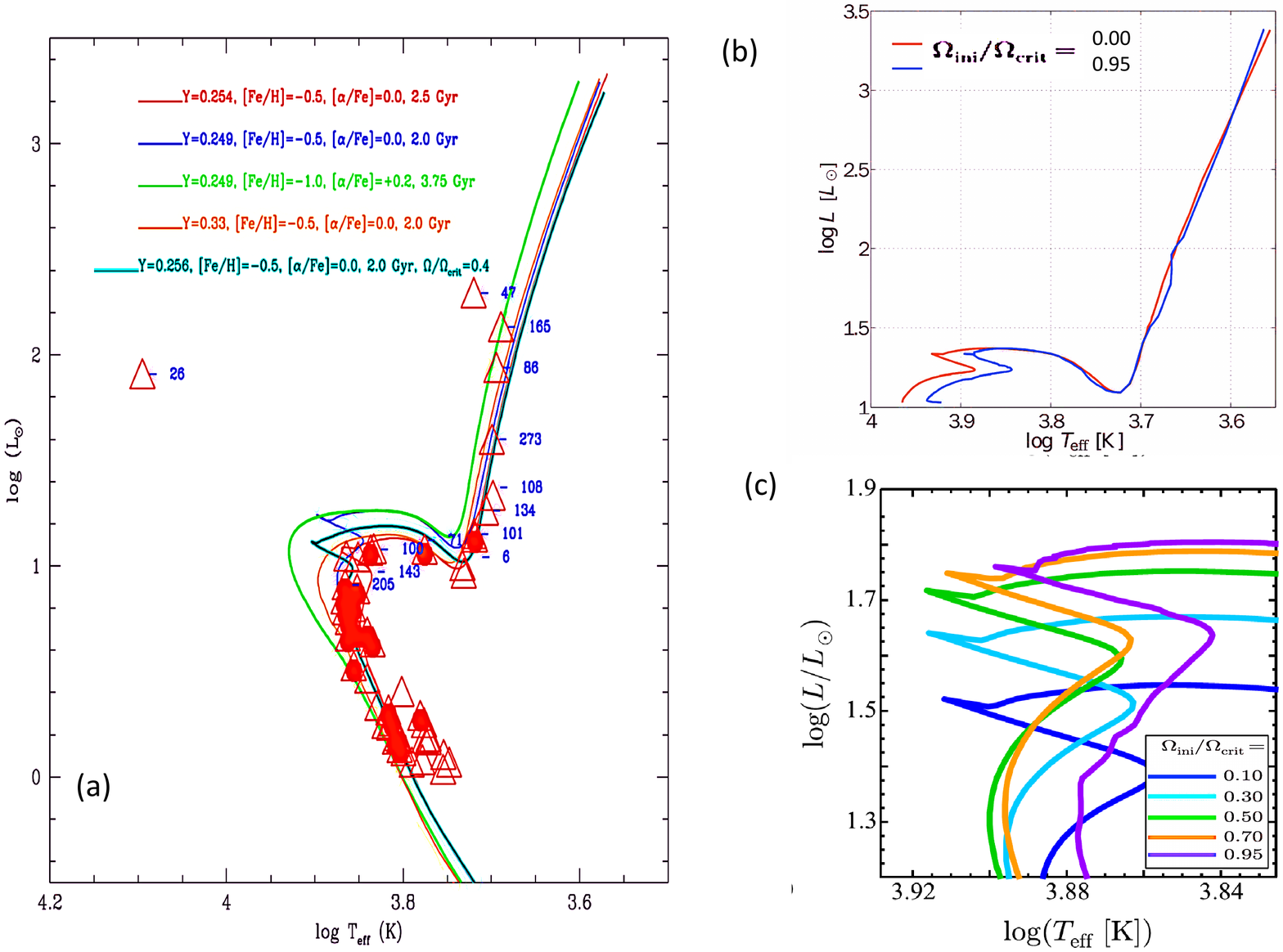}
\hspace{-0.725in}
\includegraphics[width=0.6\textwidth]{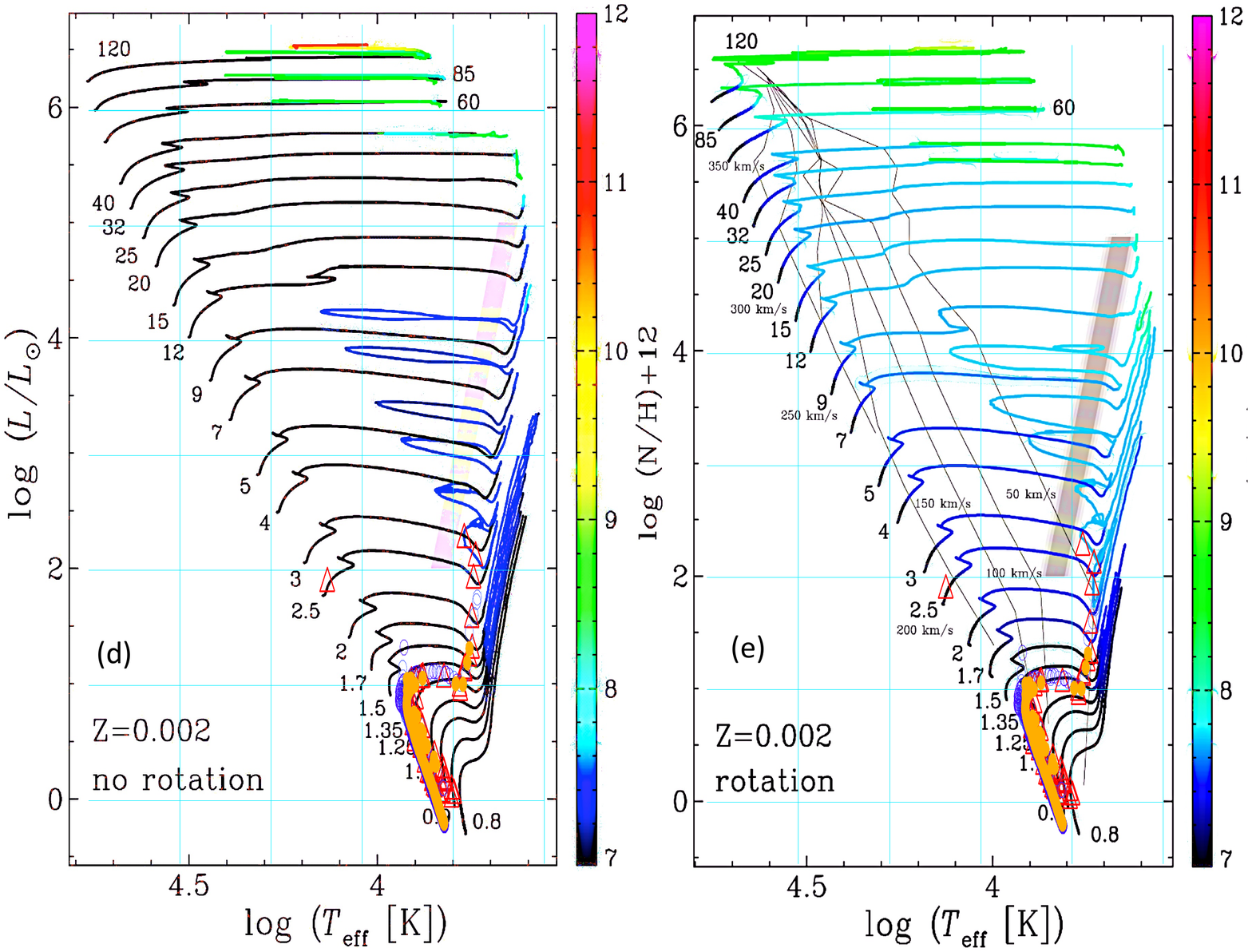}
\caption{\rm (a) \rm Translating the CMD to the $\log L(L_\odot)\; vs. \; \log T_{eff}$ plane for the D08 and D16/MIST models, the cyan/black track is the best mean fit to the MSTO  stars from F11 (red filled circles). We show the Dartmouth models for Seg~3's mass range from the MS to the SGB is $M=0.7-1.5\; M_\odot$, with the 49 brighter probable members shown as open red triangles. We number the 14 stars from Figure~10, and these always appear in the same order in the rest of Figure~12. \rm (b) \rm From \citet{li14,geo13}, Geneva models for a model-cluster of the same approximate age and metallicity with the full range of rotation rates, mass tracks for $M=1.7\; M_\odot$ are shown. \rm (c) \rm From  \citet{geo13}, mass tracks for $M=2.0\; M_\odot$ are shown for a range of rotation rates. \rm (d) \rm Geneva-model clusters with 250 stars with $Z=0.002$, with non-rotating models with $\log (Age)=9.45$ and 9.5 (light blue open circles  and gold circles, respectively), and \rm (e) \rm the same Segue 3 stars as open red triangles, with $\Omega /\Omega_{crit}=0.568$ models with $\log (Age)=9.45$, on the mass-tracks with rotation  from \citet{geo13}.
}
\label{fig:f12}
\end{figure}

\newpage 
 \section{Age vs. Metallicity}
 
 The implied split at the MSTO is more likely due to a difference in rotation rates and binarity, not age, where this conclusion is supported by the narrow SGB.  Seg~3's age and metallicity resemble a very-sparse, disrupted version of the SMC cluster, NGC 1651 \citep{li14}, and other comparisons would be the outer-LMC clusters, KMHK 1751 \& 1754 \citep{pia16}.

\begin{figure}
\includegraphics[width=1.0\textwidth]{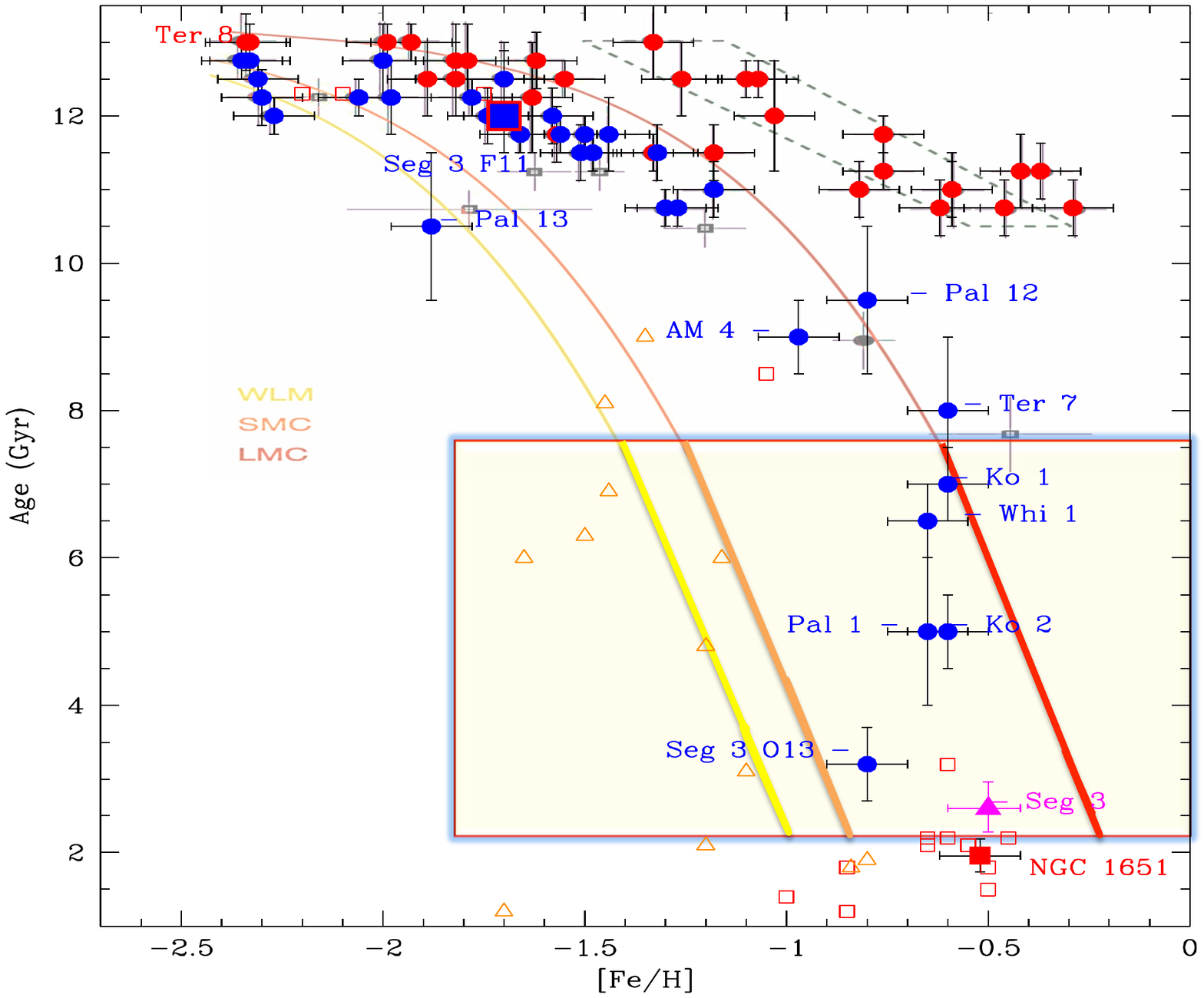}
\caption{A combination of the \citet{van13} (uniform) data on 55 MW GCs,  identified by red ($R_C<8 \; kpc$) and blue ($R_C>8\; kpc$)  --  separating clusters by their distance from the galactic center ($R_C$). The
newer halo globulars discussed in O13 are added to the HST GC, with Ko 1 \& 2 from \citet[which they identified as OCs]{pau14}. The red open squares are LMC clusters and the orange open triangles are the SMC
clusters used by \citet{pia02}. The F11 data point is shown as a blue filled square with a red border. NGC 1651 is shown as a red filled square.
The  \citet[and references therein]{lea13a, lea13b} models for the AMRs for the WLM, SMC, and LMC dwarf galaxies are shown (extrapolated in the lemon-tinted window) and six newer halo clusters are noted as black squares. A representation of the age-metallicity relationship  of the MW bulge GCs is shown as the dashed region. NGC 1651 \citep{li14} is located just below the Seg~3 Padova result, which belongs to the LMC.
}
\label{fig:f13}
\end{figure}

It is possible, but unlikely, that [Fe/H] appears higher than -1 because of CN-enhancement \citep{cum17}. However, the drift towards higher Z shows up in all filters, including $(V-I)$, when we use the ASteCA code to remove user-bias.

Figure~13 compares MW GC data with the clusters in some nearby \rm dwarf galaxies \citep[and references therein]{lea13a}. \rm We show the age-metallicity relationship (AMR) for the MW GCs and \citet{lea13b} presented models for the WLM, SMC, and LMC dwarf galaxies are shown (and extrapolated in the yellow window inset). Some newer halo clusters are denoted as black squares. A representation of the AMR of the MW bulge GCs is shown as the dashed region. We combined the \citet{van13} data on 55 MW GCs,  identified by red ($R_C<8 \; kpc$) and blue ($R_C>8\; kpc$)  --  separating clusters by their distance from the galactic center ($R_C$). The \citet{har96} data is more comprehensive, but the \citet{van13} is more uniform. The newer halo globulars discussed in O13 are added to the HST GC, with Ko 1 \& 2 from \citet[which they identified as OCs]{pau14}. The red open squares are LMC clusters and the orange open triangles are the SMC clusters used by \citet{pia02}. The F11 data point is shown as a blue square with a red border. This study's result  is shown as a magenta triangle for the Padova/ASteCA fit. Our results indicate that Segue 3  resembles the LMC clusters, and we support O13's claim that it is the youngest globular-like cluster in the MW, although spectroscopy is needed to confirm its nature, if we are to rule out that it could be an old, sparse OC. In either case, its location in the outer halo and youth argue against it being a cluster native to our Galaxy. 

\citet{all60} noted  the $\alpha$--element enhancement in metal-poor stars compared to the solar value more than fifty years ago, and \citet{wal62} decisively found excesses of Mg, Si, Ca, and Ti, relative to Fe.  Typically, globular cluster stars show $[\alpha/Fe]$ of around +0.4, but non-cluster halo stars show more scatter in this parameter, where \citet[and references therein]{fel13} gives a recent review including MW disk(s), bulge, and halo. At low-metallicity, [Fe/H] does not necessarily scale linearly with $[Ca/H]$ or $[Ca/Fe]$ \citep{bat98}. Tests with the D08 and D16 models showed that the Seg~3 population is not significantly $\alpha$-enhanced. The stellar populations from the LMC and Sgr dSph have a different enrichment ``knee" \citep{ven04}, the metallicity where chemical enrichment of the environment changes from SN Type II to Ia.

The cluster Palomar 1 \citep{sar07,sak11} fits the AMR for the LMC in Figure~13.
The AMR  for GCs  is quite different for the MW Halo and bulge, but Seg~3 does not even
fit in with unusual clusters which were likely acquired from non-MW sources.

  When  \citet{pau14} calculated the tidal radius of Ko 1 \& 2, they used $r_t=R_{GC}(2(M_{Cl}/M_{MW}))^{1\over 3}$, which yields $r_t=24\pm 5$~pc considering mass and distance uncertainties. Figure~1's King model for the the highest mass for Seg~3, from the ASteCA runs, gives $r_t\approx 20\pm 10$~pc. The uncertainty in the King model fit results from fitting a sparse-cluster radial density profile, where the cluster is not spherical.  

 \begin{deluxetable}{lcccccccl}
\tablecaption{Final Results from All Models}
\tablehead{\colhead{Models}& \colhead{Method} &  \colhead{Filters} & \colhead{$\log Age$} &    \colhead{$Z$}&  \colhead{$[\alpha/Fe]$}&  \colhead{$(m-M)_0$} &  \colhead{$E(B-V)^a$} & \colhead{Notes}}  
\startdata
   Padova& ASteCA & $VI$&  $9.40 \pm 0.20$& $0.007 \pm 0.002$& 0.0& $17.36 \pm 0.03$& $0.09\pm 0.01$& O13; 1 run$^*$, VI only\\
    Padova& ASteCA & $SDSS-gr$&  $9.5 \pm 0.2$& $0.002 \pm 0.003$& 0.0& $17.37 \pm 0.03$& $0.08\pm 0.01$&  1 run$^*$\\
    Padova& ASteCA & $CT_1ugrVI$&  $9.38 \pm 0.11$& $0.006 \pm 0.002$& 0.0& $17.33 \pm 0.08$& $0.09\pm 0.01$& Mult. runs$^{**}$\\
    Padova& ASteCA &$CT_1ugrVI$&  $9.42 \pm 0.08$& $0.006 \pm 0.001$& 0.0& $17.35 \pm 0.08$& $0.09\pm 0.01$& Mult. runs$^{**}$\\
    \small D08& BASE-9& $CT_1ugr$& $9.51^{+0.14}_{-0.12}$& $0.005 \pm 0.001$& 0.0& $17.18^{+0.06}_{-0.04}$& $0.11\pm 0.01$& $\tiny Y=0.27\pm0.02{^d}{^e}$$^{**}$ \\
    \small D08& Manual$^b$& $CT_1ugr$& $9.35^{+0.10}_{-0.31}$& $0.006 \pm 0.003$& $0.14\pm 0.15$& $17.33^c$& $\small 0.09$& $\tiny Y=0.30\pm0.06{^d}{^e}$ \\
    \small  MIST& Manual$^b$& $CT_1$& $9.35^{+0.10}_{-0.35}$& $0.004 \pm 0.001$& 0.0& $17.33^c$& $0.09$& $\tiny Y=0.245+1.6Z{^d}{^f}$\\
    \rm Geneva& \rm Manual$^b$& $\rm CT_1$& $\rm 9.4\pm 0.1$& $\rm 0.002^g$& \rm 0.0& $\rm 17.33^c$& $\rm 0.09$&  $\rm Y=0.248$\\ 
     \enddata
     \tablecomments{Model control:}
     \tablenotetext{a}{Set to 0.09 or only allowed to vary 0.05--0.15.}
     \tablenotetext{b}{``Manual" $\chi^2$-fit to model grid.}
      \tablenotetext{c}{Set to best average/O13 value.}
       \tablenotetext{d}{49 stars.}
        \tablenotetext{e}{Variable helium-content.}
         \tablenotetext{f}{Y set by Z.}
          \tablenotetext{\rm g}{\rm The only metal-poor Geneva model with variable rotation used.}
         \tablenotetext{*}{``Automatic" mode, mid-range membership and cluster-radius}
         \tablenotetext{**}{~~Semi-automatic mode, varying membership stringency and cluster-size determination.}
\end{deluxetable}

\section{Summary and Conclusions}

The mean position of Segue 3 center in all filters is $R.A.=320.\degr38015$ and $Dec.=19.\degr11753$. From cluster-cleaning and background subtraction experiments, fitting a 3-parameter King model yields  a cluster radius of 
$r_{cl}=0.\degr017\pm 0.\degr007$, a core radius is $r_{c}=0.\degr003\pm 0.\degr001$, and the tidal radius is likely to be $r_t=0.\degr04 \pm 0.\degr02$.
From all runs of the ASteCA code, iterating toward agreement between the O13, Washington, and SDSS CMDs, the Padova/PARSEC12 solar-scaled models show that Segue 3 has $Z=0.006 \pm 0.002$, $\log(Age)=9.38 \pm 0.11$, $(m-M)_0=17.33 \pm 0.08$, $E(B-V)=0.09\pm 0.01$, with a binary fraction of $0.36\pm 0.12$. The mass estimate from the King models was quite uncertain, $630\pm 264 \; M_\odot$. If we only use the  runs where we remove the SDSS-identified galaxies, and use the information from F11 on radial-velocity members: $Z=0.006 \pm 0.001$, $\log(Age)=9.42 \pm 0.08$, $(m-M)_0=17.35 \pm 0.08$, $E(B-V)=0.09\pm 0.01$, with a binary fraction of $0.39\pm 0.05$, giving a cluster mass of $478\pm 56 \; M_\odot$. With $Z=0.006$ and an age of $2.6^{+0.6}_{-0.4}$ Gyr, our estimates for Segue 3 are younger and more metal-rich than O13's result, using the same set of isochrones. Converting metallicity to iron abundance: $\mathrm{[Fe/H]}\approx -0.5$ if the cluster is not $\alpha$-enhanced, and $\mathrm{[Fe/H]}<-0.5$ if $[\alpha/Fe]$ is positive. Seg~3  does not follow the AMR trend of MW-native globular clusters, resembling field stars and clusters from  the (gas-rich) WLM dIrr, SMC, and LMC.  Comparing the GC data with \citet{lea13a,lea13b} models indicates that a system with Seg~3's properties could have been formed in an LMC/SMC-like system, extrapolating their ``leaky box" models. The results of this paper are summarized in Table 6. 

\rm  We support O13's results: Seg~3 is certainly the youngest ``GC" found in the MW to date, and it was more massive in the past. Our results favor Seg~3 being a disrupting-GC and not an OC. \rm 
  When we re-analyzed the O13 data using the ASteCA code, the results for the distance modulus
are consistent with the Washington and SDSS analysis, but SDSS colors are less sensitive to $E(B-V)$ than Washington colors. However, the broader C-filter allows us to go deeper than u-band. Our analysis shows the importance of comparing clusters within the same model-grid. The D16/MIST models (though solar-scaled) are more useful for estimates of the age \rm uncertainty \rm in Washington colors because of the finer grid and the extension of the models past the helium-flash. The D08 models include a range of $\alpha$-abundances and Y-values, which are used with BASE-9 \citep{wag16a,wag16b}: for a single-population model, Seg~3 has $Y=0.27\pm 0.02$, but this might be a rotation effect making stars bluer below the MSTO.

Though unusual, Segue 3 is not unique in being a young, metal-rich  cluster in the MW outer halo (see Figure~13).  Others have been found in the MW, including those suspected to be associated with the Sgr dSph 
\citep[for Pal 12, Ter 7, and Whiting 1, respectively]{coh04,sbo05,law10}.  Other MW clusters are similarly low mass, and fall  somewhere between traditional open and globular clusters  \citep[e.g., Pal 1]{sak11} 
 or seem to be massive, old open clusters \citep[who studied Ko 1 and 2]{pau14}.  Indeed, Pal 12, Ter 7, and Pal 1 all  show low $[\alpha/Fe]$-abundances \citep{coh04,sbo05,law10} typical of dwarf galaxy stars \citep{tol09}, which  do not follow the standard MW alpha-enhancement \citep{all60,wal62}.  Tests with D08 and D16 models suggest that Seg  3 is not significantly $\alpha$-enhanced, but a firm conclusion requires 
spectroscopic follow-up.  However, there is circumstantial evidence that Seg~3 came from an accreted gas-rich system \citep[B10, F11, O13]{bel07}.

\acknowledgments

The authors thank  Julie Lutz,   Jeff Brown,  and the UW Astronomy ``Stars" discussion group for their input.  We also acknowledge financial support from the NSF, the M.J. Murdock Charitable Trust, and the Kennilworth Fund of the New York Community Trust. We thank  Gabriel Perren for modifying the ASteCA code for us (the current version is available here: http://asteca.github.io). This work has made use of BaSTI web tools. We thank the Rachel Wagner-Kaiser, TvH, DvD, DS, and ER, and the BASE-9 team for their efforts. Hughes thanks the APO 3.5-m Users Committee for input into the design of the ARCTIC imager, and for the APO staff's help with remote observing. \rm We also thank the anonymous referee for helpful suggestions that improved the flow of the manuscript. \rm

This research has made use of the NASA/IPAC Infrared Science Archive, which is operated by the Jet Propulsion Laboratory, California Institute of Technology, under contract with the National Aeronautics and Space Administration. We used data from the 3.5-m telescope's APIcam and ARCTIC imagers. We also made extensive use of the Sloan Digital Sky Survey. Funding for the Sloan Digital Sky Survey IV has been provided by
the Alfred P. Sloan Foundation, the U.S. Department of Energy Office of
Science, and the Participating Institutions. SDSS-IV acknowledges
support and resources from the Center for High-Performance Computing at
the University of Utah. The SDSS web site is www.sdss.org.

SDSS-IV is managed by the Astrophysical Research Consortium for the 
Participating Institutions of the SDSS Collaboration including the 
Brazilian Participation Group, the Carnegie Institution for Science, 
Carnegie Mellon University, the Chilean Participation Group, the French Participation Group, Harvard-Smithsonian Center for Astrophysics, 
Instituto de Astrof\'isica de Canarias, The Johns Hopkins University, 
Kavli Institute for the Physics and Mathematics of the Universe (IPMU) / 
University of Tokyo, Lawrence Berkeley National Laboratory, 
Leibniz Institut f\"ur Astrophysik Potsdam (AIP),  
Max-Planck-Institut f\"ur Astronomie (MPIA Heidelberg), 
Max-Planck-Institut f\"ur Astrophysik (MPA Garching), 
Max-Planck-Institut f\"ur Extraterrestrische Physik (MPE), 
National Astronomical Observatories of China, New Mexico State University, 
New York University, University of Notre Dame, 
Observat\'ario Nacional / MCTI, The Ohio State University, 
Pennsylvania State University, Shanghai Astronomical Observatory, 
United Kingdom Participation Group,
Universidad Nacional Aut\'onoma de M\'exico, University of Arizona, 
University of Colorado Boulder, University of Oxford, University of Portsmouth, 
University of Utah, University of Virginia, University of Washington, University of Wisconsin, 
Vanderbilt University, and Yale University.



\vspace{5mm}
\facilities{APO: 3.5m}

\software{IRAF, Python, ASteCA, BASE-9, SYCLIST}

\end{document}